\documentclass[12pt]{article}
\pdfoutput=1

\usepackage{graphicx,psfrag,epsf,xcolor}
\usepackage{amsmath,amssymb,amsfonts}
\usepackage{array}
\usepackage{cite}
\usepackage{rotating}
\usepackage{slashed, cancel}
\usepackage{scalerel,stackengine}
\usepackage{float}
\usepackage{subcaption}
\usepackage{caption}
\definecolor{cBlue}{RGB}{0,110,191}
\usepackage[pdftitle={Next-to-leading power two-loop soft functions for the Drell-Yan process at threshold},hypertexnames=true]{hyperref}
\hypersetup{bookmarksnumbered,colorlinks,
    linkcolor={black},
    citecolor={cBlue},
    urlcolor={cBlue}}
\numberwithin{equation}{section}

\newcommand{\be}{\begin{equation}}
\newcommand{\ee}{\end{equation}}
\newcommand{\bea}{\begin{eqnarray}}
\newcommand{\eea}{\end{eqnarray}}
\newcommand{\bi}{\begin{itemize}}
\newcommand{\ei}{\end{itemize}}
\newcommand{\ben}{\begin{enumerate}}
\newcommand{\een}{\end{enumerate}}
\newcommand{\bt}{\begin{tabular}}
\newcommand{\et}{\end{tabular}}

\newcommand{\nn}{\nonumber}

\definecolor{darkgreen}{rgb}{0.0,0.6,0.0}
\definecolor{cPurple}{RGB}{93,35,125}

\begin{document}
\allowdisplaybreaks

\begin{titlepage}
		
\begin{flushright}
{\small
IPPP/21/06\\
CERN-TH-2021-108\\
\today
}
\end{flushright}
		
\vskip0.7cm
\begin{center}
{\Large \bf\boldmath
Next-to-leading power two-loop soft functions \\[0.2cm]
for the Drell-Yan process at threshold}
\end{center}
		
\vspace{0.5cm}
\begin{center}
{\sc Alessandro Broggio},$^{a,b}$
{\sc Sebastian Jaskiewicz},$^{c}$ 
and {\sc Leonardo Vernazza}$^{d,e}$\\[6mm]

{\it $^a$ Universit\`a degli Studi di Milano-Bicocca \\
Piazza della Scienza 3, I--20126 Milano, Italy\\[0.2cm]
}
{\it $^b$ INFN, Sezione di Milano-Bicocca \\
Piazza della Scienza 3, I--20126 Milano, Italy\\[0.2cm]}
{\it $^c$ Institute for Particle Physics Phenomenology, Durham University\\
Durham DH1 3LE, United Kingdom 
 \\[0.2cm]
}
{\it $^d$ Theoretical Physics Department, CERN, CH-1211 Geneva 23, Switzerland
 \\[0.2cm]
}
{\it $^e$ Dipartimento di Fisica Teorica, 
Universit\`a di Torino \\
and INFN, Sezione di Torino, Via P. Giuria 1, 
I-10125 Torino, Italy 
}
\end{center}
		
\vspace{0.4cm}
\begin{abstract}
\vskip0.2cm\noindent
We calculate the generalized soft functions at~$\mathcal{O}(\alpha_s^2)$ at next-to-leading power accuracy for the Drell-Yan process at threshold.
The operator definitions of these objects contain explicit insertions of soft gauge and matter fields, giving rise to a dependence on additional convolution variables with respect to the leading power result.
These soft functions constitute the last missing ingredient for the validation of the bare factorization theorem to NNLO accuracy.
We carry out the calculations by reducing the soft squared amplitudes into a set of canonical master integrals and we employ the method of differential equations to evaluate them.
We retain the exact $d$-dimensional dependence of the convolution variables at the integration boundaries in order to regulate the fixed-order convolution integrals.
After combining our soft functions with the relevant collinear functions, we perform checks of the results at the cross-section level against the literature and expansion-by-regions calculations, at NNLO and partly at N$^3$LO, finding agreement.
\end{abstract}
\end{titlepage}

\section{Introduction}
\label{sec:introduction}

Substantial progress has recently been achieved in the study of subleading power corrections in elementary scattering processes within the soft-collinear effective theory (SCET) \cite{Bauer:2000yr,Bauer:2001yt,Beneke:2002ph,Beneke:2002ni} framework. Next-to-leading power (NLP)
leading logarithmic (LL) contributions to the partonic cross sections were resummed to all orders in perturbation theory using renormalization group (RG) techniques for event shape observables \cite{Moult:2018jjd}, and for the parton diagonal channels of colour singlet production processes, such as $q \bar{q} \to \gamma^*$ \cite{Beneke:2018gvs} and $gg \to H$ \cite{Beneke:2019mua}, near the kinematic threshold.\footnote{NLP LL results were obtained using diagrammatic techniques in \cite{Bahjat-Abbas:2019fqa} and compared to SCET in \cite{vanBeekveld:2021hhv}. } Progress beyond the current state of the art has been hindered by the ubiquitous appearance of endpoint divergent convolution integrals \cite{Beneke:2008pi,Benzke:2010js,Beneke:2019oqx} in subleading power factorization formulas both at LL in the parton non-diagonal channels, and at next-to-leading logarithmic (NLL) accuracy for the diagonal channels.
These divergences prohibit a straightforward application of the standard RG methods to perform resummation.
Solutions to this problem at LL accuracy have been found in particular cases by employing consistency relations \cite{Moult:2019uhz}, refactorization conditions \cite{Liu:2019oav} and a combination of operator refactorization and consistency relations \cite{Beneke:2020ibj}.
At NLL accuracy a solution has been obtained for the $h\to \gamma \gamma$ decay mediated by light-quarks using diagrammatic methods \cite{Anastasiou:2020vkr} and within the SCET$_{{\rm{II}}}$ framework \cite{Liu:2020wbn,Liu:2020tzd}.
However, a universal solution to these problems is not currently known.

By employing the basis of subleading $N$-jet operators constructed in  \cite{Beneke:2017ztn,Beneke:2018rbh,Beneke:2019kgv},\footnote{A power suppressed operator basis in the label formulation of SCET can be found in~\cite{Marcantonini:2008qn,Kolodrubetz:2016uim,Feige:2017zci,Moult:2017rpl}.}
the bare factorization theorem for the $q\bar{q}$-initiated Drell-Yan (DY) process in the threshold region\footnote{An extension of the standard threshold limit to include full collinear dynamics was studied in \cite{Lustermans:2019cau}.} was derived at general subleading powers in \cite{Beneke:2019oqx}.
In particular, the next-to-leading power factorization formula is proportional to the leading power (LP) hard function and to convolutions between the generalized soft functions and their associated collinear functions, the latter calculated to $\mathcal{O}(\alpha_s)$ in \cite{Beneke:2019oqx}.
Formally, this result is only valid when the soft and collinear functions are evaluated in exact $d$-dimensions before evaluating the convolution integrals. However, after convolution, it is safe to expand in the $d\to 4$ limit.

Currently, the two-loop generalized soft functions are the only missing ingredients which are required to validate the DY factorization theorem \cite{Beneke:2019oqx} up to NNLO at NLP accuracy.
The aim of the present article is to fill this gap by providing the calculation of the generalized soft functions at $\mathcal{O}(\alpha_s^2)$ which is carried out while retaining the relevant $d$-dimensional dependence of the results.
We reduce the squared amplitudes to master integrals (MIs) by employing the program \texttt{LiteRed} \cite{Lee:2012cn,Lee:2013mka} and we calculate the MIs  using the method of differential equations and the transformation to the canonical basis \cite{Henn:2013pwa}.
The results for the soft functions are validated at the cross-section level, \emph{after} convolution with the collinear functions, by directly comparing to results obtained by means of the expansion-by-regions and diagrammatic methods~\cite{Laenen:2010uz,Bonocore:2016awd,Bahjat-Abbas:2018hpv}. We also find agreement with the NLP contribution of the NNLO result in \cite{Hamberg:1990np}. To achieve this, we sum the contributions to the cross-section due to the soft functions  with the remaining  NNLO contributions calculated in \cite{Beneke:2019oqx} and collected in App.~\ref{appNNLOterms}. In the last step, we carry out the remaining UV renormalization and remove the initial state collinear singularities at cross section level.

To the best of our knowledge, this is the first time that soft functions at NLP are evaluated to $\mathcal{O}(\alpha^2_s)$.
However, in the present work, we do not analyze the UV renormalization and the RG evolution of the soft functions since any such procedure requires an expansion around $d\to 4$ which leads to the appearance of divergent convolutions integrals preventing naive renormalization attempts.
Our intention is to provide more information about the higher order structure of these soft functions which could give a hint towards the solution of the divergent convolution problem, at least in the present case.
However, it should be noted that at NLP accuracy, calculations and studies of the renormalization and evolution properties of the soft function needed for the $h \to \gamma \gamma$ decay process were recently presented at $\mathcal{O}(\alpha_s)$ in \cite{Liu:2020eqe,Bodwin:2021cpx}.
In the context of the $q_T$-subtraction method, calculations at fixed-order accuracy, which required the evaluation of several new integrals, have been recently carried out at NLP in QCD \cite{Oleari:2020wvt} without separating the different regions. 

The paper is organized as follows. In Sec.~\ref{sec:NLPfactorizationReview}
we review the structure of the factorized cross section which is one of the main results of~\cite{Beneke:2019oqx}.
In Sec.~\ref{sec:TwoLoopSoft} we describe in detail the calculation of the two-loop soft functions. In particular, we discuss the evaluation of the canonical master integrals using the differential equation method.
In Sec.~\ref{sec:Validation} we calculate the convolution integrals of the soft functions with the corresponding collinear functions~\cite{Beneke:2019oqx} which allows us to carry out a series of checks at the cross-section level against the literature, \cite{Hamberg:1990np,Laenen:2010uz,Bonocore:2016awd}, and against expansion-by-regions calculations, both at NNLO and available results at N$^3$LO \cite{Bahjat-Abbas:2018hpv}.  In App.~\ref{app:collinearfns} we list the analytic results for the collinear functions calculated in~\cite{Beneke:2019oqx}, and in App.~\ref{sec:matrixelements} we provide expressions for the relevant two-parton matrix elements used in this calculation.  Useful cross-section level results from \cite{Beneke:2019oqx} are collected in App.~\ref{appNNLOterms}.   Finally, App.~\ref{AppAP} contains the expressions for the relevant Altarelli-Parisi splitting functions.

\section{Factorization near threshold}
\label{sec:NLPfactorizationReview}
In this section, we review the structure of the NLP factorization theorem for the DY process in the threshold region \cite{Beneke:2019oqx}, and we remind the reader of  the operatorial definitions of the NLP soft functions which are relevant for this work.

We consider the parton diagonal channel of the DY process, $q\bar q\to\gamma^* [\to\ell\bar{\ell}\,]+X$ in the kinematic region $z=Q^2/\hat{s}\to 1$, where $\hat{s}=x_ax_b \,s$ is the partonic centre-of-mass energy squared, $x_a,x_b$ are the momentum fractions of the partons inside the incoming hadrons and $Q^2$ is the invariant mass squared of the lepton pair.
Up to NLP in the threshold expansion, the cross-section differential in $Q^2$ assumes the following form
\begin{equation}
\frac{d\sigma_{\rm DY}}{dQ^2} = 
\frac{4\pi\alpha_{\rm em}^2}{3 N_c Q^4}
\sum_{a,b} \int_0^1 dx_a dx_b\,f_{a/A}(x_a)f_{b/B}(x_b)
\Big( 
\hat{\sigma}^{\,{\rm{LP}}}_{ab}(z)\,
+ {\hat{\sigma}^{\,{\rm{NLP}}}_{ab}(z)} + \mathcal{O}(\lambda^4)
\Big) + \mathcal{O}\left(\frac{\Lambda}{Q}\right),
\label{eq:dsigsq2}
\end{equation}
where $f_{a/A}(x_a)$ and $f_{b/B}(x_b)$ are the usual parton distribution functions (PDFs), and $\hat{\sigma}(z)$ with superscripts LP and NLP are the leading power and the next-to-leading power partonic cross sections, respectively.
Since we only focus on the $q\bar{q}$-channel, we omit the indices $a,b$ in the following.
The LP partonic cross section factorizes into a product of two functions 
\begin{equation}
\hat{\sigma}_{\rm{LP}}(z) = H(Q^2) \,Q \, S_{\rm DY}(Q(1-z))\,,
\label{eq:LPfactv2}
\end{equation}
the hard function $H(Q^2)$ and the soft function
\begin{equation}
S_{\rm DY}(\Omega) = \int \frac{dx^0}{4\pi}\,e^{i \Omega\, x^0 /2}\,
\widetilde{S}_{0}\left(x^0 \right)\,.
\label{eq:LPsoftfn}
\end{equation}
The LP position-space soft function is a vacuum matrix element of soft Wilson lines\footnote{This object and its relation to other LP soft functions has been recently investigated in~\cite{Falcioni:2019nxk}.} \cite{Korchemsky:1993uz}
\begin{eqnarray}\label{eq:LPsoft}
\widetilde{S}_{0}(x ) &=& 
\frac{1}{N_c}{\rm{Tr}}\, \langle 0| \bar{\mathbf{ T}} \left[ 
Y_{+}^\dagger(x)Y_{-}(x)  \right]  
{\mathbf{ T}}
\left[ Y_{-}^\dagger(0) Y_{+}(0) \right]
|0 \rangle\,,
\end{eqnarray}
where
\begin{eqnarray}
	Y_{\pm}\left(x\right)&=&\mathbf{P}
	\exp\left[ig_s\int_{-\infty}^{0}ds\,n_{\mp}
	A_{s}\left(x+sn_{\mp}\right)\right].
\end{eqnarray}
We now turn our attention to the NLP part of the factorization formula, which is understood
to be formally valid only in $d$-dimensions, before renormalization.
To facilitate comparison with literature, we define the quantity $\Delta$ related to the partonic cross section as follows 
\begin{eqnarray}
 \label{Delta}
\Delta(z) = \frac{1}{(1-\epsilon)} \frac{\hat{\sigma}(z)}{z}\,. 
\end{eqnarray}
The NLP partonic cross section receives contributions from power corrections to the phase-space, the so-called \emph{kinematic} corrections, and from insertions of subleading power Lagrangian terms in time-ordered product operators, referred to as \emph{dynamical} corrections. We have 
\begin{eqnarray}\label{kinPlusdyn}
\Delta_{\rm{NLP}}(z)= \Delta^{dyn }_{{\rm{NLP}} }(z) + \Delta^{kin }_{{\rm{NLP}} }(z)\,.
\end{eqnarray}
The $\Delta^{kin }_{{\rm{NLP}} }(z)$ term has been presented in Eq.~(5.11) of~\cite{Beneke:2019oqx} at NNLO. In this work, we focus on the calculation of the generalized soft functions which appear in the factorization formula in the $\Delta^{dyn }_{{\rm{NLP}} }(z)$ contribution. The result for $\Delta^{dyn }_{{\rm{NLP}} }(z)$ takes the following form \cite{Beneke:2019oqx} 
\begin{eqnarray}
\label{eq:3.24}
\Delta^{dyn }_{{\rm{NLP}} }(z)&=&- \frac{2}{(1-\epsilon)} \,  
Q \left[ \left(\frac{\slashed{n}_-}{4}
\right) {\gamma}_{\perp\rho}  \left(\frac{\slashed{n}_+}{4}
\right) \gamma^{\rho}_{\perp} \right]_{\beta\gamma}
\nonumber  \\ 
&&  \hspace{0cm} \times 
\, \int d(n_+p)\,C^{\,{A0,A0}\,}(n_+p, x_bn_-{p}_B  ) \,
C^{*A0A0}\left(\,x_an_+p_A,\,x_b{n_-p_B}\right)
\nonumber \\ 
&&  \hspace{0cm} \times \,\sum^5_{i=1}
\,\int \left\{d\omega_j\right\}
{J}_{i,\gamma\beta}\left(n_+p,x_a n_+p_A; 
\left\{\omega_j\right\} \right) \, 
{S}_{i}(\Omega; \left\{\omega_j\right\} ) 
+\rm{h.c.}\,,\quad
\end{eqnarray}
where $\Omega=Q(1-z)$. In the equation above, $C^{\,{A0,A0}\,}$ is the hard matching coefficient of the LP SCET current for the DY process. The $J_i$ are the collinear functions and the $S_i$ represent the generalized soft functions in momentum space, defined as
\be\label{3.82eq}
{S}_{i}(\Omega; \left\{\omega_j\right\} ) = 
\int \frac{dx^0}{4\pi} \,  e^{i \Omega\, x^0/2}  
\int \bigg\{\frac{dz_{j-}}{2\pi} \bigg\} \, e^{-i\omega_j{z_{j-}}}
{S}_{i}(x_0; \left\{z_{j-}\right\} ) \,,
\ee
in terms of the position-space multi-local soft functions, ${S}_{i}(x_0; \left\{z_{j-}\right\} )$.
At NLP these are given by
\begin{eqnarray}
\label{eq:3.23}
&&{S}_{1}(x^0; z_- )\,=\,
\frac{1}{N_c}\, {\rm{Tr}} \langle 0| \bar{\mathbf{ T}} \left[ Y_{+}^\dagger(x^0)
Y_{-}(x^0)  \right] {\mathbf{ T}}\left(
\left[ Y_{-}^\dagger(0) Y_{+}(0) \right]
\frac{i\partial_{\perp}^{\nu}}{in_-\partial}
\mathcal{B}^{+}_{\nu_{\perp}}\left(z_{-}\right)
\right)|0 \rangle\,, \quad\qquad
\\[2ex] \label{eq:3.24c}
&&{S}^{}_{2;\mu\nu}(x^0; z_- )\,=\, \frac{1}{N_c}\,
{\rm{Tr}}\,\langle 0|  \bar{\mathbf{ T}} \left[ Y_{+}^\dagger(x^0) Y_{-}(x^0) 
\right]   \nn\\ 
&&\hspace{2.5cm} \times\, {\mathbf{ T}}\left(
\left[ Y_{-}^\dagger(0) Y_{+}(0) \right]
\frac{1}{(in_-\partial)}
\big[  \mathcal{B}^+_{\mu_\perp}(z_-)
,\mathcal{B}^+_{\nu_\perp}(z_-)\big] 
\right)|0 \rangle\,,
\\[2ex]
&&{S}^{}_{3}(x^0; z_-)\,=\,\frac{1}{N_c} \,{\rm{Tr}}\,\langle 0|  
\bar{\mathbf{ T}} \left[ Y_{+}^\dagger(x^0) Y_{-}(x^0) \right]
\nn \\&& \hspace{2.5cm}\times\, {\mathbf{ T}}\left(
\left[ Y_{-}^\dagger(0) Y_{+}(0) \right]
\frac{1}{(in_-\partial)^2}\left[ 
\mathcal{B}^{+\,\mu_\perp}(z_{-}),
\left[in_-\partial\mathcal{B}^+_{\mu_\perp}(z_{-})
\right]  \right] 
\right)|0 \rangle\,,
\label{eq:3.26}
\\[2ex]\label{eq:3.26b}
&&{S}^{AB}_{4;\mu\nu,bf}(x^0; z_{1-},z_{2-} )\,=\,
\frac{1}{N_c} \,{\rm{Tr}}\, \langle 0| 
\bar{\mathbf{ T}} \left[ Y_{+}^\dagger(x^0) Y_{-}(x^0) \right]_{ba}
\nn\\&&\hspace{2.5cm} \times \,{\mathbf{ T}}\left(
\left[ Y_{-}^\dagger(0) Y_{+}(0) \right]_{af}
\mathcal{B}^{+A}_{\mu_\perp}(z_{1-})
\mathcal{B}^{+B}_{\nu_\perp}(z_{2-})
\right)|0 \rangle\,,
\\[2ex]
\label{eq:3.27}
&&{S}_{5;bfgh,\sigma\lambda}(x^0; z_{1-},z_{2-} )\,=\,
\frac{1}{N_c} \,\langle 0|  
\bar{\mathbf{ T}} \left[ Y_{+}^\dagger(x^0) Y_{-}(x^0) \right]_{ba}
\nn\\ &&\hspace{2.5cm}\times\,
{\mathbf{ T}}\left(
\left[ Y_{-}^\dagger(0) Y_{+}(0) \right]_{af}
\frac{g_s^2}{(in_-\partial_{z_1})(in_-\partial_{z_2})}
{q}_{+\sigma g}(z_{1-})
\bar{q}_{+\lambda h}(z_{2-})
\right)|0 \rangle\,.
\end{eqnarray}
In the above definitions, $\mu, \nu$ are the Lorentz indices, $\sigma,\lambda$ are the Dirac indices, and $A,B$ and $a,b,f,g,h$ are adjoint and fundamental colour indices, respectively. The $\mathcal{B}_{\pm}$($ {q}_{+}$) field is a soft building block formed by a soft covariant derivative (soft quark field) and soft Wilson lines 
\begin{eqnarray} 
	\mathcal{B}_{\pm}^{\mu} &=& Y_{\pm}^{\dagger}
	\left[ i\,D^{\mu}_s\,Y_{\pm}\right] \,,
\quad\quad \quad
	{q}^{\pm}   =  Y_{\pm}^{\dagger}\, q_s \,.
\end{eqnarray}
The soft functions in Eqs.~\eqref{eq:3.23} -- \eqref{eq:3.27} are the fundamental objects of interest in this work. Thus far, only partial results for these objects have been reported in the literature. The $\mathcal{O}(\alpha_s)$ result for $S_1$, expanded in $\epsilon$, was given in~\cite{Beneke:2018gvs}, and with complete $d$-dimensional dependence in~\cite{Beneke:2019oqx}.
At $\mathcal{O}(\alpha_s^2)$, results for virtual-real soft diagrams have been presented in~\cite{Beneke:2019oqx}.
It was found that only the $S_1$ soft function receives such contributions.
In this article, we complete the calculations of the bare soft functions at~$\mathcal{O}(\alpha_s^2)$ by considering all the diagrams with two-real soft parton emissions.

A quick inspection of the soft functions in \eqref{eq:3.23} -- \eqref{eq:3.27} reveals that $S_1$ and $S_3$ are conveniently defined as scalar objects. The remaining three functions, $S_2$, $S_4$ and $S_5$, contain instead  a non-trivial dependence on Lorentz, Dirac, and adjoint and fundamental colour indices.
These indices are contracted with the corresponding indices carried by the respective collinear functions, which are reported for convenience in App.~\ref{app:collinearfns}.
We prefer to work with scalar objects and, for this reason, we absorb the colour, spin and Lorentz structures of the multi-local collinear functions into their corresponding soft functions $S_4$ and $S_5$.
For example, making use of~\eqref{eq:J4tree}, the part of the factorization formula at $\mathcal{O}(\alpha^2_s)$ which depends on $S_4$ is given by, \footnote{We indicate the perturbative order in $\alpha^n_s$ with the corresponding superscript ${(n)}$.}
\begin{eqnarray}
\label{eq:3.24S4}
\Delta^{dyn (2)}_{{\rm{NLP}}, S_4 }(z)&=&  4  \,  
Q\, H^{(0)}(Q^2)
\int  d\omega_1 d\omega_2\, 
J^{\mu\nu,AB\,(0)}_{4; fb }
\left(x_an_+p_A;\omega_1,\omega_2\right) \, 
{S}^{AB (2)}_{4;\mu\nu,bf}(\Omega; \omega_{1},\omega_{2} ). 
\,\, \quad
\end{eqnarray}
Focusing on the collinear and soft functions, we now redefine
\begin{eqnarray}\label{eq:redefinition}
J^{\mu\nu,AB\,(0)}_{4;   fb }
\left(n_+p;\omega_1,\omega_2\right) 
{S}^{AB (2)}_{4;\mu\nu,bf}(\Omega; \omega_{1},\omega_{2} )  \equiv  \underbrace{\left(-\frac{1}{n_+p}\right)}_{= J_4^{(0)}} \,
{S}^{(2)}_{4}(\Omega; \omega_{1},\omega_{2} ),\quad
\end{eqnarray}
such that the new $J_4^{(0)}$ and $S^{(2)}_4$ are scalar functions.
The ${S}_{5;bfgh,\sigma\lambda}$ soft function in Eq.~\eqref{eq:3.27} is redefined in an analogous way by factoring out the same scalar collinear function such that $J^{(0)}_5=J^{(0)}_4$.  Additionally, in this case, the spin structures in the first line of \eqref{eq:3.24} are also absorbed into the soft function.

The $S_{2;\mu\nu}$ soft function is anti-symmetric under the exchange of the Lorentz indices~\mbox{$\mu,\nu$}. Since this is a vacuum matrix element, it must be proportional to the epsilon tensor $\epsilon^{\mu\nu}_{\perp}$ which is the only anti-symmetric structure which can carry two transverse Lorentz indices. However, its appearance is forbidden due to parity conservation of~QCD. Indeed, we checked by direct calculation that $S_{2;\mu\nu}$ vanishes at~$\mathcal{O}(\alpha_s^2)$.

\section{Two-loop soft functions}
\label{sec:TwoLoopSoft}

We proceed with the main focus of this work by providing the calculation details of the double real emission corrections to the soft functions defined in~\eqref{eq:3.23} -- \eqref{eq:3.27}. Techniques for solving integrals which appear in calculations of LP soft functions at NNLO have been developed over the years and several examples exist in the SCET literature. In particular, results for the exclusive soft function relevant for small transverse momentum resummation in DY was obtained in~\cite{Li:2011zp}, the soft function for the production of an electroweak boson at large transverse momentum was computed in \cite{Becher:2012za}, and the calculation of the soft function relevant for boosted top-quark pair production was presented in \cite{Ferroglia:2012uy}.
However, these methods are insufficient for the NNLO calculation of the NLP soft functions containing dependence on additional convolution variables. Therefore, we apply more advanced techniques developed for fixed order calculations. Similar methods were used in  \cite{Wang:2018vgu}  to calculate the NNLO soft function for top quark pair production at threshold. The strategy is straightforward: we first obtain the squared amplitudes at $\mathcal{O}(\alpha_s^2)$. Subsequently we use \texttt{LiteRed} \cite{Lee:2012cn,Lee:2013mka} to reduce such expressions to a linear combination of master integrals (MIs), and finally, we compute the necessary MIs by employing the differential equation method. Each of these steps is expanded upon in the following sections.

\subsection{Reduction to master integrals}
\label{sec:reduction}
\begin{figure}
\begin{centering}	
\subcaptionbox{}{
  \includegraphics[width=0.32\textwidth]{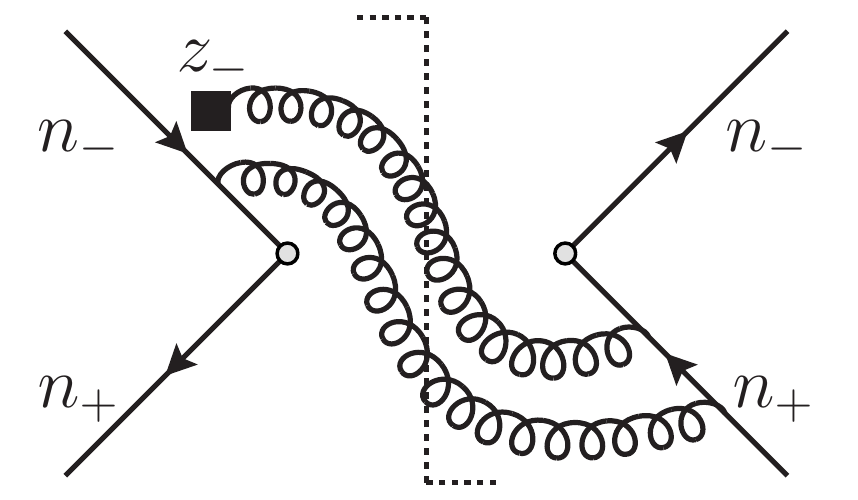} }
\subcaptionbox{}{
 \includegraphics[width=0.32\textwidth]{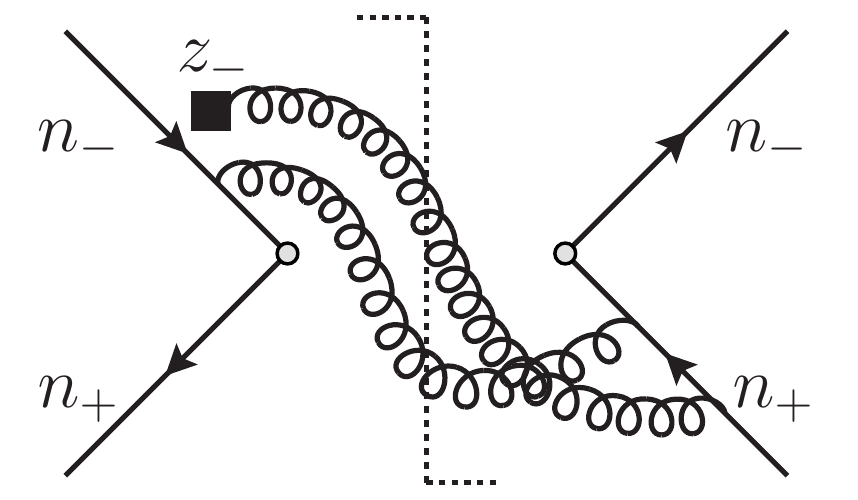}}
\subcaptionbox{}{
\includegraphics[width=0.32\textwidth]{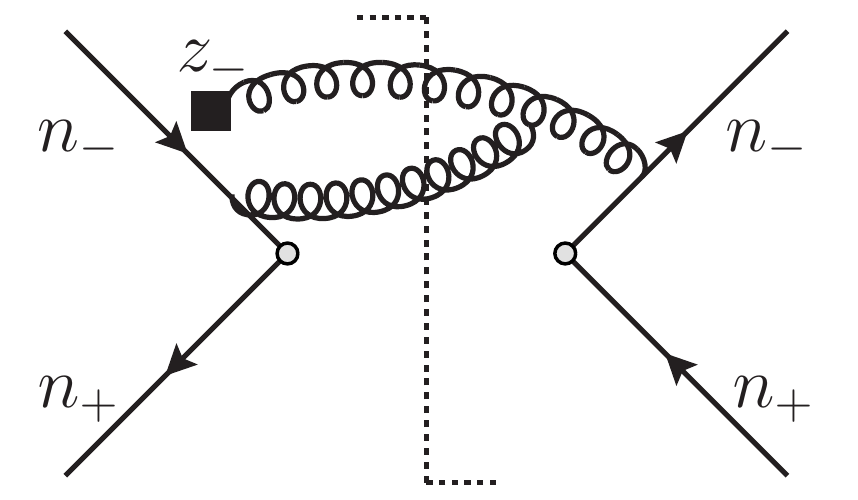}}
\subcaptionbox{}{\includegraphics[width=0.32\textwidth]{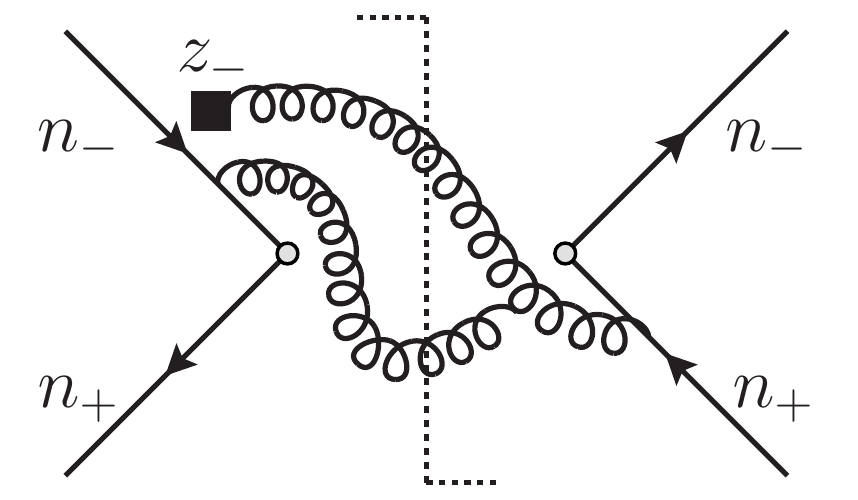}}
\subcaptionbox{}{\includegraphics[width=0.32\textwidth]{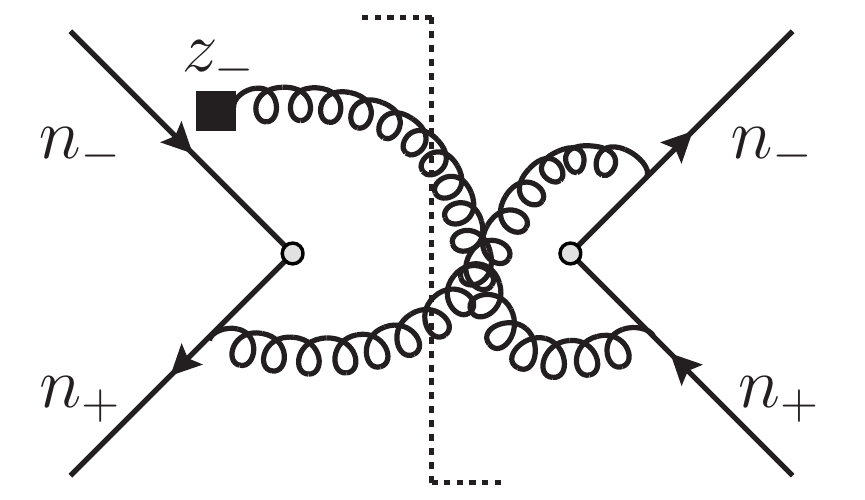}}
\subcaptionbox{}{\includegraphics[width=0.32\textwidth]{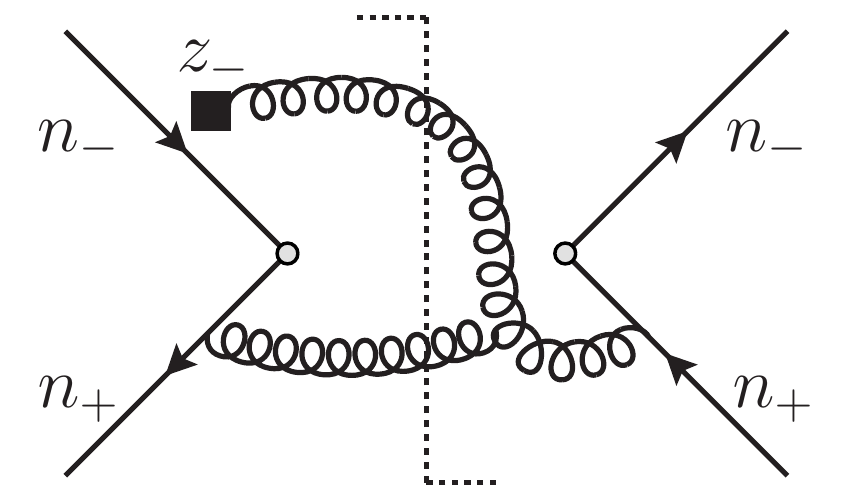}}
\subcaptionbox{}{\includegraphics[width=0.32\textwidth]{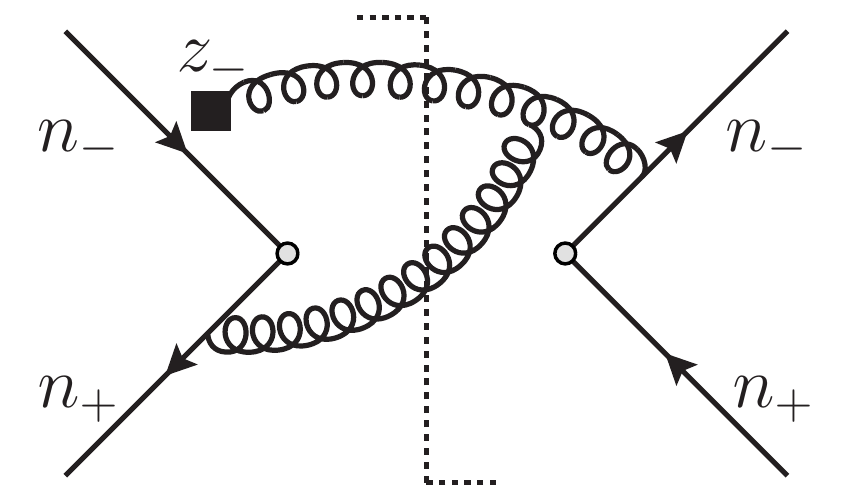}}
\subcaptionbox{}{\includegraphics[width=0.32\textwidth]{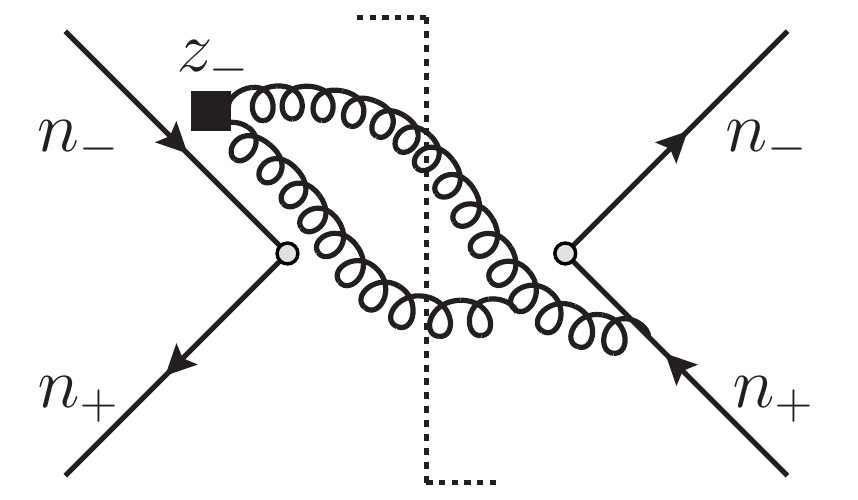}}
\subcaptionbox{}{\includegraphics[width=0.32\textwidth]{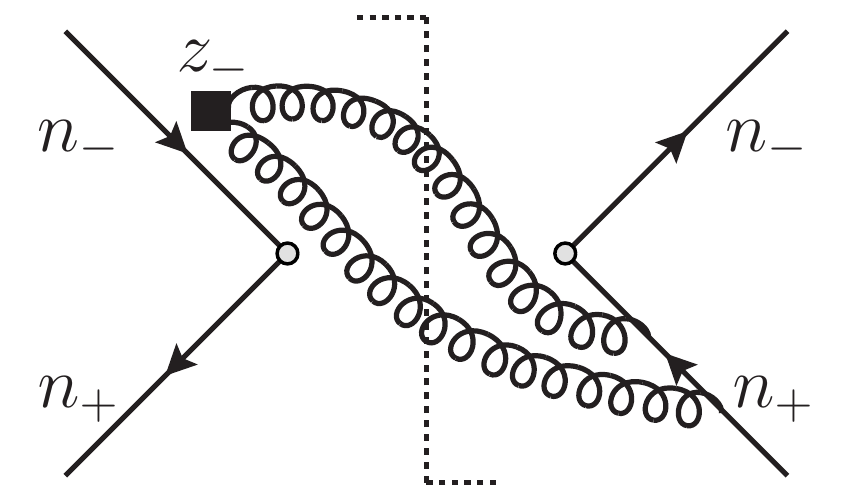}}
\subcaptionbox{}{\includegraphics[width=0.32\textwidth]{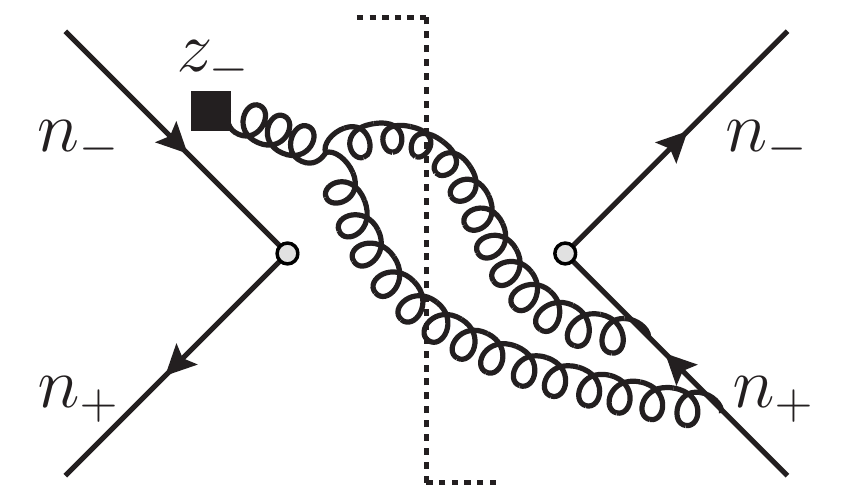}}
\subcaptionbox{}{\includegraphics[width=0.32\textwidth]{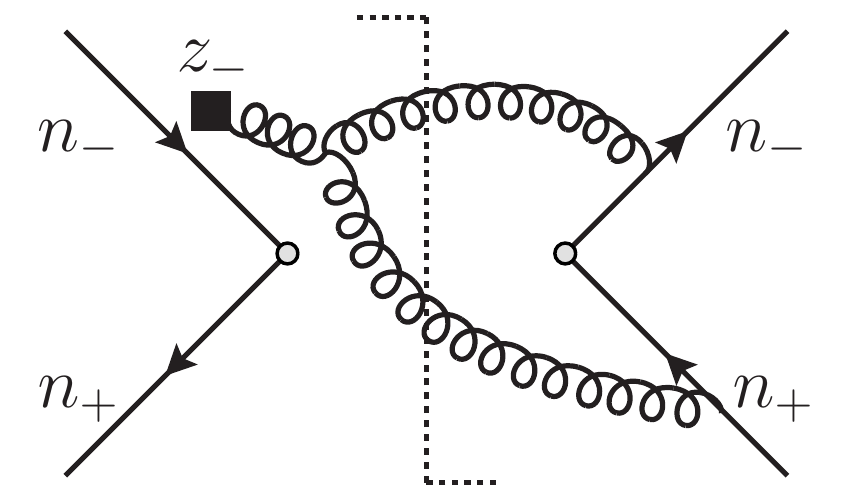}}
\subcaptionbox{}{\includegraphics[width=0.32\textwidth]{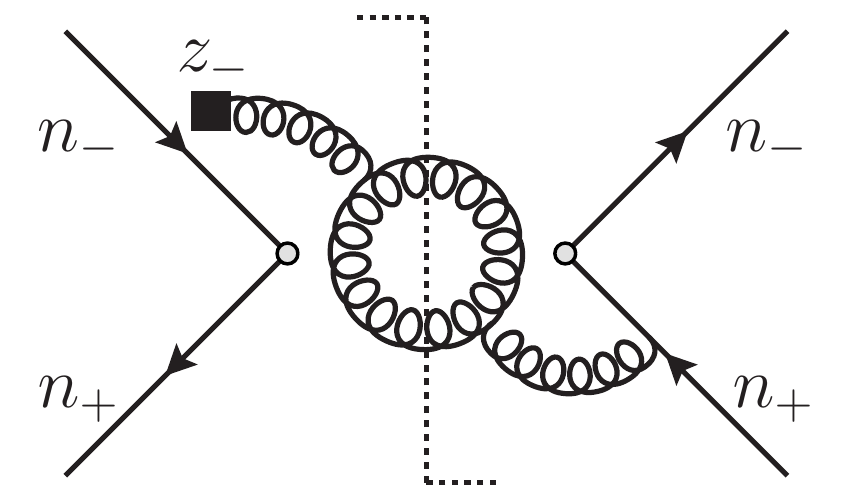}}
\subcaptionbox{}{\includegraphics[width=0.32\textwidth]{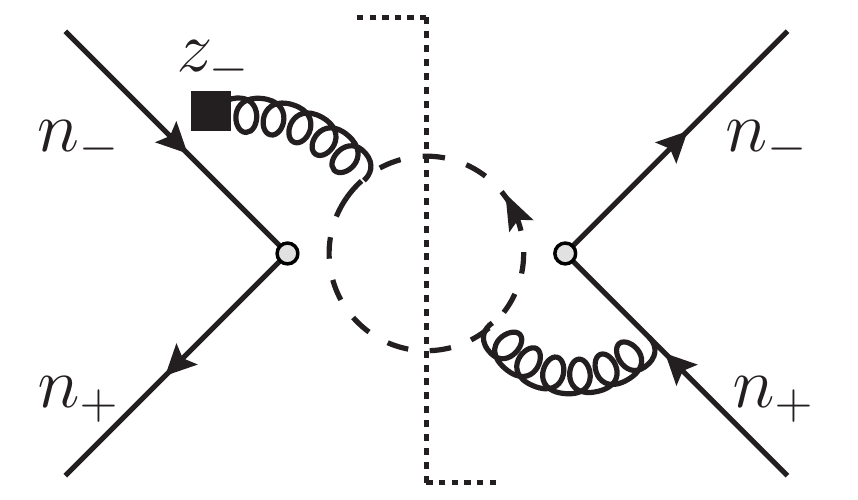}}
\subcaptionbox{}{\includegraphics[width=0.32\textwidth]{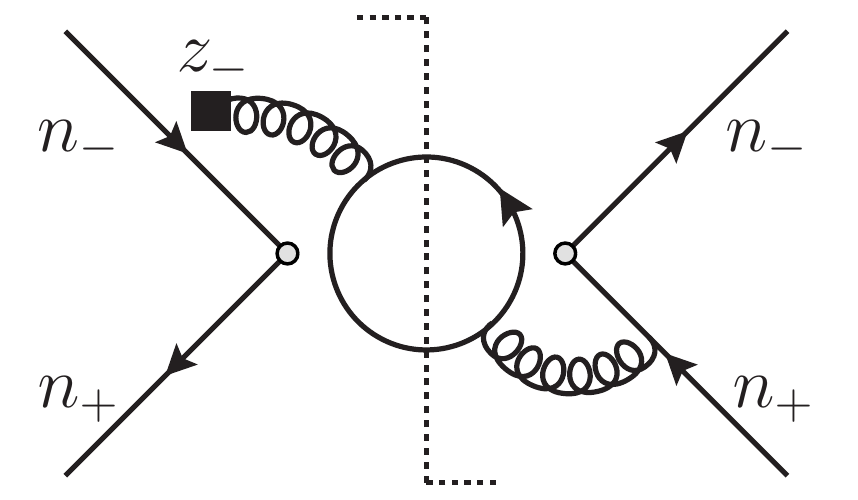}}
 \par\end{centering}	\caption{\label{fig:S1} Diagrams contributing to the $S_1$ soft function.
The part to the left (right) of the cut corresponds to the time-ordered (anti-time-ordered) part of the diagram, and lines labeled by $n_{\pm}$ with in (out)-going arrow correspond to soft Wilson lines $Y_{\mp}$($Y_{\mp}^{\dagger}$).  The filled square in this figure stands for the soft covariant derivative and the Wilson lines contained in  $\frac{i\partial_{\perp\mu}}{in_-\partial}\mathcal{B}_+^{\mu}=\frac{i\partial_{\perp\mu}}{in_-\partial}  Y^{\dagger}_+[iD^{\mu}_{s}Y_+]$.
}\end{figure}

\begin{figure}	\begin{centering}	\subcaptionbox{}{\includegraphics[width=0.32\textwidth]{averagepictures/2lspaper_v2_diag9.pdf}}
\subcaptionbox{}{\includegraphics[width=0.32\textwidth]{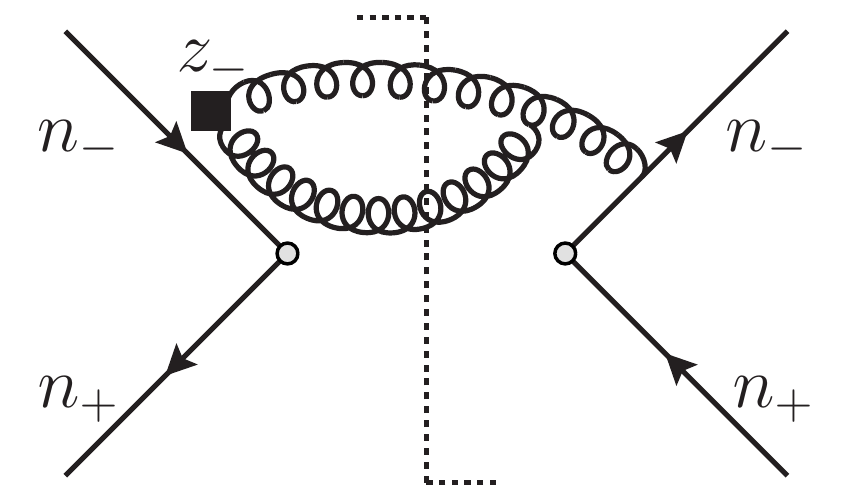}}
\subcaptionbox{}{\includegraphics[width=0.32\textwidth]{averagepictures/2lspaper_v2_diag8.pdf}}
\par\end{centering}	\caption{\label{fig:S3diagrams} Diagrams contributing to the $S_3$ soft function. Same conventions as in Figure~\ref{fig:S1} are used. Here, the filled square stands for the soft covariant derivatives and the Wilson lines contained in  $\frac{1}{(in_-\partial)^2}\left[ 
\mathcal{B}^{+\,\mu_\perp} ,
\left[in_-\partial\mathcal{B}^+_{\mu_\perp} 
\right]  \right]$.
}\end{figure}

\begin{figure}	\begin{centering}	\subcaptionbox{}{\includegraphics[width=0.32\textwidth]{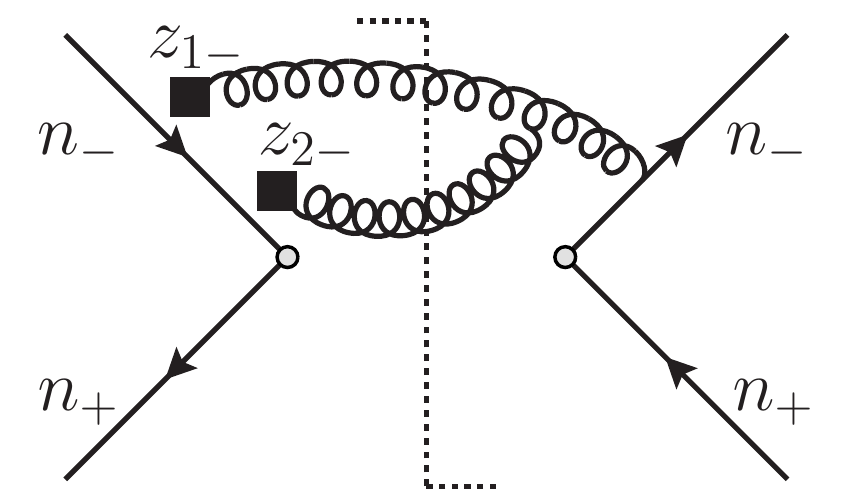}}
\subcaptionbox{}{\includegraphics[width=0.32\textwidth]{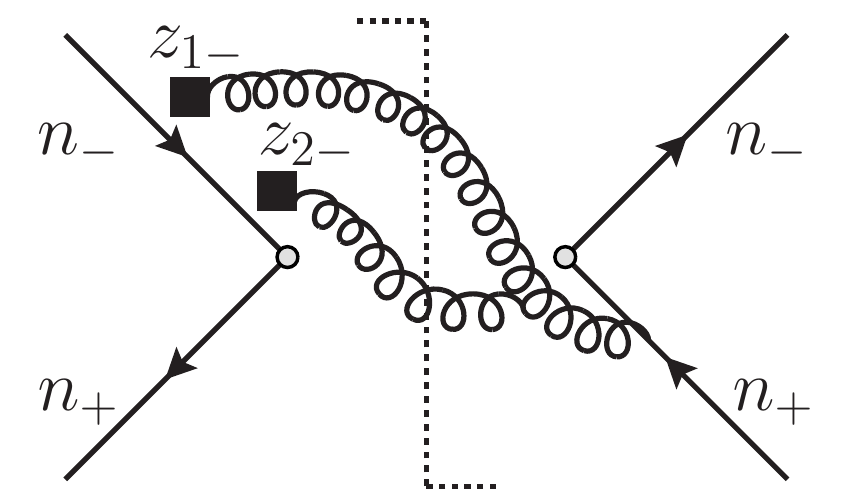}}
\par\end{centering}	\caption{\label{fig:S4diagrams} Diagrams contributing to the $S_4$ soft function.
Same conventions as in Figure~\ref{fig:S1} are used. The filled squares stand for the soft covariant derivative and the Wilson lines contained in $\mathcal{B}^{+\, \mu_\perp}(z_{1-})
\mathcal{B}^{+}_{\mu_\perp}(z_{2-})$.
}\end{figure}

\begin{figure}	\begin{centering}	\subcaptionbox{}{\includegraphics[width=0.32\textwidth]{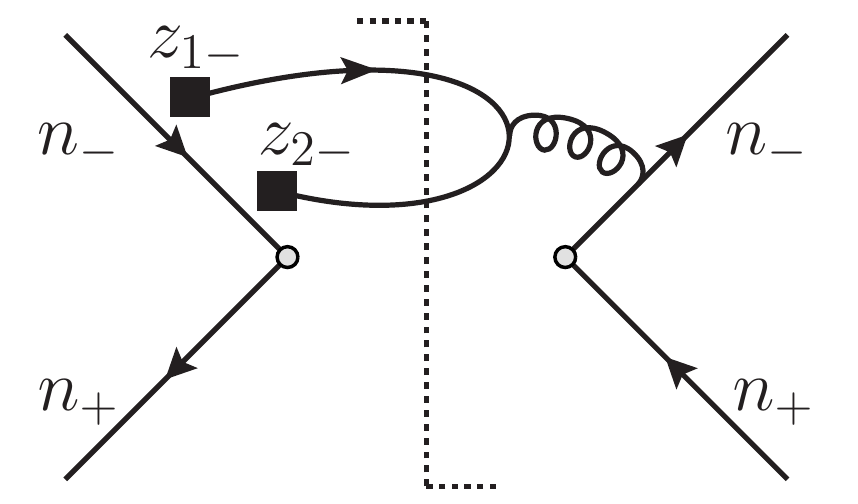}}
\subcaptionbox{}{\includegraphics[width=0.32\textwidth]{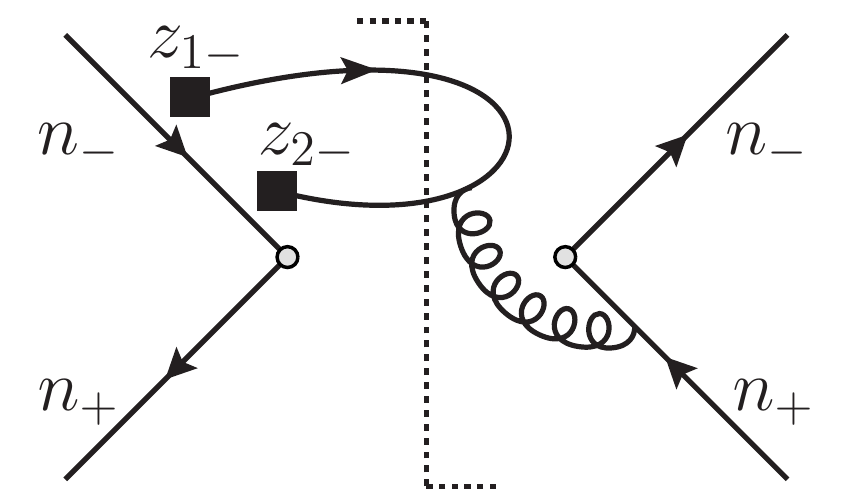}}
\par\end{centering}	\caption{\label{fig:S5diagrams} Diagrams contributing to the $S_5$ soft function.
Same conventions as in Figure~\ref{fig:S1} are used. The filled squares in this figure stand for the soft covariant derivative and the Wilson lines contained in ${q}_{+ }(z_{1-})\, \bar{q}_{+ }(z_{2-})$.
}\end{figure}

The two-loop expressions for the soft functions $S_1$ and $S_3$ are directly obtained from their matrix element definitions.
The soft functions $S_4$ and $S_5$ are computed after the redefinition made in \eqref{eq:redefinition} by employing the NLP Feynman rules given in \cite{Beneke:2018rbh}.
The expressions for the squared matrix elements of all the soft functions are collected in the \texttt{ancillary.pdf} file and they correspond to the diagrams in Figures \ref{fig:S1} -- \ref{fig:S5diagrams}. We use $k_1$ and $k_2$ to label the momenta of the partons crossing the cut.

\subsubsection{Topologies}
\label{sec:topologies}

The calculation of the double real emission corrections to the soft functions includes two types of phase space constraints.
The first constrains the total energy radiated into the final state, enforced by the $\delta(\Omega-2E_X)$ condition, where $E_X$ is the total radiation energy.
The second type instead constrains specific light-cone components of the soft parton momenta.
Indeed, the NLP soft functions are also differential in $\omega$, or $\omega_1$ and $\omega_2$, which are the convolution variables that connect the soft functions to their corresponding collinear functions, as prescribed by~\eqref{eq:3.24}.
We find that up to $\mathcal{O}(\alpha_s^2)$, three different constraints of the second type are possible. Namely, the integrands of the soft functions depend on $\delta(\omega-n_-k_1)$, \footnote{In principle the $\delta(\omega-n_-k_2)$ contribution can also appear, but it is mapped back to the structure $\delta(\omega-n_-k_1)$ by relabelling the momenta.} or $\delta(\omega-n_-k_1-n_-k_2)$ or $\delta(\omega_1-n_-k_1)\, \delta(\omega_2-n_-k_2)$. These constraints, along with the on-shell cut propagators conditions $\delta(k_1^2)$ and $\delta(k_2^2)$, are set in the \texttt{LiteRed} program.
We now define the auxiliary topologies, $\mathcal{A}$, $\mathcal{B}$, and $\mathcal{C}$, which implement the $\delta(\omega-n_-k_1)$ constraint and only differ among themselves by the choice of one propagator.
Topology $\mathcal{A}$ is defined by the following set of seven propagators
\begin{align}\label{topologySA}
P_1&=(k_1+k_2)^2,\quad P_2=n_+k_2,\quad P_3=n_-k_2,\,  \\\nonumber
P_4&=k_1^2,\quad  P_5=k_2^2,\quad P_6=\big(\Omega - n_-k_1 - n_-k_2 -n_+k_1 - n_+k_2 \big),\quad P_7 = \big(\omega-n_-k_1\big)\, ,
\end{align}
where the last four propagators are cut propagators.
This means that
\begin{align}
\frac{1}{P_4}\to \delta(k^2_1)=\frac{1}{2 \pi i}\bigg[\frac{1}{k_1^2 + i0^+} - \frac{1}{k_1^2 - i0^+}\bigg] \, ,
\end{align}
and equivalent relations hold for $P_5$, $P_6$ and $P_7$.
Topology $\mathcal{B}$ is obtained starting from the list of propagators in \eqref{topologySA} and replacing the single propagator \mbox{$P_3\to n_-(k_1+k_2)$}.
Similarly, topology $\mathcal{C}$ requires the substitution \mbox{$P_3 \to n_+(k_1+k_2)$}.
The integrals which appear in our calculations are written as
\begin{align}
\hat{I}_{\mathcal{T}}(\alpha_1,\alpha_2,\alpha_3,\alpha_4, \alpha_5,\alpha_6,\alpha_7) = (4\pi)^4\bigg( \frac{e^{\gamma_E} \mu^2}{4 \pi}\bigg)^{2 \epsilon}\int \frac{d^d k_1}{(2 \pi)^{d-1}} \, \frac{d^d k_2}{(2 \pi)^{d-1}} \prod_{i=1}^7\, \frac{1}{P^{\alpha_i}_i}\,,
\end{align}
where the index $\mathcal{T}$ indicates the specific topology $\mathcal{T}\in \{\mathcal{A},\mathcal{B},\mathcal{C},\ldots\}$, and can be expressed as a linear combination of the independent MIs.
It turns out that all the MIs for the topologies $\mathcal{A}$ and $\mathcal{C}$ are a subset of the MIs for the topology $\mathcal{B}$. This is not surprising since the three topologies share most of the propagators. In particular, we find the following five MIs for topology $\mathcal{B}$
\begin{align}
&\hat{I}_1(\Omega,\omega)\equiv \hat{I}_{\mathcal{B}}(0, 0, 0, 1, 1, 1, 1),\quad & \hat{I}_2(\Omega,\omega)\equiv \hat{I}_{\mathcal{B}}(0, 0, 1, 1, 1, 1, 1) , \nonumber \\
& \hat{I}_3(\Omega,\omega)\equiv \hat{I}_{\mathcal{B}}(1, 0, 0, 1, 1, 1, 1),\quad & \hat{I}_4(\Omega,\omega)\equiv \hat{I}_{\mathcal{B}}(1, 1, 0, 1, 1, 1, 1), \nonumber  \\
& \hat{I}_5(\Omega,\omega) \equiv \hat{I}_{\mathcal{B}}(1, 1, 1, 1, 1, 1, 1)\, ,\label{eq:MISSB}
\end{align}
where the integral $\hat{I}_1(\Omega,\omega)$ represents the phase space integral.
The squared matrix elements with the $\delta(\omega-n_-k_1-n_-k_2)$ constraint require four additional topologies to be reduced.
The topology $\mathcal{D}$ is defined by the list of propagators
\begin{align}\label{topologySD}
P_1&=(k_1+k_2)^2,\quad P_2=n_+k_2,\quad P_3=n_-k_2,\quad P_4=k_1^2, \\\nonumber
  P_5&=k_2^2,\quad P_6=\big(\Omega - n_-k_1 - n_-k_2 -n_+k_1 - n_+k_2 \big),\quad P_7 = \big(\omega-n_-k_1-n_-k_2\big)\, ,
\end{align}
where $P_4$ to $P_7$ are cut. Topology $\mathcal{E}$ is obtained from \eqref{topologySD} by replacing \mbox{$P_3 \to n_-k_1$}, topology $\mathcal{F}$ by substituting \mbox{$P_2 \to n_+k_1$}, and the topology $\mathcal{G}$ by exchanging both \mbox{$P_2 \to n_+k_1$} and \mbox{$P_3 \to n_-k_1$}. After reduction, we find that only two additional MIs appear for the set of topologies which implement the constraint $\delta(\omega-n_-k_1-n_-k_2)$:
\begin{align}\label{eq:I6I7}
&\hat{I}_6(\Omega,\omega)\equiv \hat{I}_{\mathcal{E}}(0, 0, 0, 1, 1, 1, 1),\quad & \hat{I}_7(\Omega,\omega)\equiv \hat{I}_{\mathcal{E}}(1, 1, 1, 1, 1, 1, 1)\, .
\end{align}
Two additional topologies are needed to reduce the integral expressions with a double constraint given by $\delta(\omega_1-n_-k_1)\, \delta(\omega_2-n_-k_2)$. We define the $\mathcal{H}$ topology as
\begin{align}\label{topologySH}
P_1&=(k_1+k_2)^2,\quad P_2=n_+k_2,\quad P_3=k_1^2\quad P_4=k_2^2 , \\\nonumber
  \quad P_5&=\big(\Omega - n_-k_1 - n_-k_2 -n_+k_1 - n_+k_2 \big),\quad P_6 = \big(\omega_1-n_-k_1\big)\, \quad P_7 = \big(\omega_2-n_-k_2\big),
\end{align}
where only the first two propagators of the list remain uncut. The topology $\mathcal{I}$ is related to $\mathcal{H}$ by a replacement of the second propagator $P_2 \to n_+k_1$. Only one MI is found for these last topologies 
\begin{align}\label{eq:I8}
&\hat{I}_8(\Omega,\omega_1,\omega_2)\equiv \hat{I}_{\mathcal{H}}(0, 0, 1, 1, 1, 1, 1) .
\end{align}
In total we find eight new MIs which need to be computed to evaluate the $\mathcal{O}(\alpha^2_s)$ corrections to the NLP soft functions.

\subsubsection{Results after reduction}
\label{sec:reductionresults}
The integrals belonging to each of the topologies defined above can be reduced by employing the program \texttt{LiteRed}. In this subsection we present the results for the soft functions in terms of linear combinations of MIs. The soft function $S_1$ carries an additional $2r0v$ superscript since it is the only one that also receives a virtual-real (superscript $1r1v$) contribution. In the following expressions we omit for simplicity the $\Omega$ and $\omega$ ($\omega_1$,$\omega_2$) dependence in the MIs and we find
 \begin{eqnarray}\label{eq:S1wk1reduced}
S^{(2)2r0v}_1(\Omega,\omega)  &=& 
\frac{\alpha_s^2}{(4\pi)^2}
C_F^2\frac{8 \left(2-9 \epsilon +9 \epsilon^2\right)}{\epsilon^2\,\omega  \, (\Omega -\omega )^2}\,\hat{I}_1 \nonumber\\&&\hspace{-2.3cm}+\frac{\alpha_s^2}{(4\pi)^2}C_FC_A \Bigg[\frac{ (2-3 \epsilon ) \left(-4 \Omega  +\epsilon\,     (  \omega +19 \Omega )+4\epsilon^2 \left( \omega  - 7 \Omega  \right)   -16\epsilon^3(\omega-\Omega) \right)}{\epsilon^2(1-2 \epsilon )\,\omega\,  \Omega   \,  (\Omega -\omega )^2}\,\hat{I}_1 
\nonumber\\&&\hspace{-0.8cm} -\frac{(1-4\epsilon^2)  }{\epsilon \,\omega \, \Omega   } \hat{I}_2 
+\frac{(3 \Omega -10\epsilon \, \Omega +16\epsilon^2 (\omega+\Omega))  }{2(1-2\epsilon) \,\omega \, \Omega   } \hat{I}_3  +  
\frac{  (\Omega -3 \omega )}{2 \omega }\,\hat{I}_4 \,
\nonumber \\ && \hspace{-0.7cm}
 +   \,\Omega\,\hat{I}_5 \,
+  \frac{(9-20 \epsilon +12 \epsilon^2 - 2 \epsilon^3)}{\epsilon^2\,  (3-2 \epsilon )\omega^2 (\Omega -\omega )}
 \hat{I}_6 
 + (\Omega -\omega ) \hat{I}_7 
\Bigg]\,
\nonumber \\ && \hspace{-2.3cm}
- \frac{\alpha_s^2}{(4\pi)^2}C_F\,n_f\, 
\frac{4 (1 - \epsilon )^2}{\epsilon \,  (3 -2 \epsilon )\omega^2 (\Omega -\omega )}\,
 \hat{I}_6 ,
\end{eqnarray}
where $\epsilon = (4-d)/2$.
The $S_3$ soft function has the following form in terms of MIs
\begin{eqnarray}\label{eq:S3reduced}
 S^{(2)}_3(\Omega,\omega) &=&\frac{\alpha_s^2}{(4\pi)^2}C_FC_A\frac{2 (1-\epsilon)}{  (3-2\epsilon)\omega^2(\Omega-\omega)}
\hat{I}_6 \,.
\end{eqnarray}
$S_4$ ans $S_5$ originate from double insertions of $\mathcal{O}(\lambda)$ power suppressed Lagrangian contributions.
We obtain
\begin{eqnarray}\label{eq:S4reduced}
 S^{(2)}_4(\Omega,\omega_1,\omega_2)
 =
-\frac{\alpha_s^2}{(4\pi)^2}C_FC_A
\frac{2        (1-\epsilon ) \omega_2( \omega_1 - \omega_2 )  }{( \omega_1 + \omega_2 )^4 (\Omega - \omega_1 - \omega_2 )}
\hat{I}_8 \,,
\end{eqnarray}
and
\begin{eqnarray}
 S^{(2)}_5(\Omega,\omega_1,\omega_2)
 =\frac{\alpha_s^2}{(4\pi)^2}\left(C_F^2-\frac{1}{2}C_FC_A\right)
 \frac{8(-1+\epsilon)\omega_2}{( \omega_1 + \omega_2 )^3 (\Omega - \omega_1 - \omega_2 )}
 \hat{I}_8 .
\end{eqnarray}

\subsection{Master integrals}
\label{sec:masterintegrals}

We begin by describing the calculation of the MIs for the topology $\mathcal{B}$ given in \eqref{eq:MISSB}. Starting from those expressions, it is convenient to make the variable change $\omega \to r \,\Omega$,
and redefine the MIs by factoring out their mass dimensions in $\Omega$
\begin{align}\label{eq:intredef}
&I^\prime_1(r) = \frac{1}{\Omega^2} \bigg(\frac{\Omega}{\mu}\bigg)^{4 \epsilon} \hat{I}_1(\Omega,r) ,\quad & I^\prime_2(r) = \frac{1}{\Omega} \bigg(\frac{\Omega}{\mu}\bigg)^{4 \epsilon} \hat{I}_2(\Omega,r), \nonumber \\
&I^\prime_3(r) = \bigg(\frac{\Omega}{\mu}\bigg)^{4 \epsilon} \hat{I}_3(\Omega, r),\quad & I^\prime_4(r) = \Omega  \bigg(\frac{\Omega}{\mu}\bigg)^{4 \epsilon} \hat{I}_4(\Omega,r), \nonumber \\
& I^\prime_5(r) = \Omega^2  \bigg(\frac{\Omega}{\mu}\bigg)^{4 \epsilon} \hat{I}_5(\Omega, r)\, ,
\end{align}
where the prime integrals only depend on the variable $r$.
We use \texttt{Canonica} \cite{Meyer:2017joq} to guide us in finding the canonical basis of MIs \cite{Henn:2013pwa}. In particular, this is achieved by the following transformations 
\begin{align}\label{eq:noncan}
I^\prime_1(r) &= \frac{2 (1-r)^2}{2 - 9 \epsilon + 9 \epsilon^2} I_1(r),\nonumber \\
I^\prime_2(r) & = \frac{2(r-1)}{1-5 \epsilon+6 \epsilon^2} I_1(r) - \frac{1}{\epsilon (1-2 \epsilon)} I_2(r), \nonumber\\ 
I^\prime_3(r) &= \frac{1}{\epsilon^2} I_3(r),\nonumber\\
I^\prime_4(r) &= -\frac{1}{\epsilon^2 (1-r)} I_4(r),\nonumber\\
I^\prime_5(r) &= \frac{1}{\epsilon^2 r} I_2(r) - \frac{1}{\epsilon^2 r} I_3(r)-\frac{1+r}{2 \epsilon^2  (1-r) r} I_4(r) + \frac{1}{\epsilon^2 r} I_5(r)\, ,
\end{align}
where the canonical integrals are the ones without the prime.
The system of differential equations for the vector of integrals $\vec{I}(r) \equiv \big\{I_1(r), I_2(r), I_3(r), I_4(r), I_5(r) \big\}$ is given by
\begin{align}\label{eq:diffeq}
\frac{d \vec{I}(r)}{d r} = \epsilon \, A(r)\cdot \vec{I}(r) \, ,
\end{align}
where
\begin{align}\label{eq:amatrix}
A(r) = \begin{bmatrix}\,\,\,
-\frac{1}{r}+\frac{3}{1-r} & \hspace{0.3cm}0 &\hspace{0.3cm} 0 &\hspace{0.3cm} 0 &\hspace{0.3cm} 0\\ \\
\frac{2}{r} &\hspace{0.3cm} -\frac{2}{r} &\hspace{0.3cm} 0\hspace{0.3cm} &\hspace{0.3cm} 0 &\hspace{0.3cm} 0\\ \\
\frac{2}{r} &\hspace{0.3cm} 0 &\hspace{0.3cm}-\frac{2}{r} &\hspace{0.3cm} 0 &\hspace{0.3cm} 0 \\ \\
\frac{2}{r}&\hspace{0.3cm} 0 &\hspace{0.3cm}\frac{2}{r} &\hspace{0.3cm} \frac{4}{1-r} &\hspace{0.3cm} 0 \\ \\
\frac{1}{r} &\hspace{0.3cm} 0 &\hspace{0.3cm} \frac{1}{r} &\hspace{0.3cm} \frac{1}{r} &\hspace{0.3cm} -\frac{2}{r}\,\,\,\,\,\,
\end{bmatrix}\, .
\end{align}
We notice from the structure of $A(r)$ that the integral $I_2(r)$ only couples to $I_1(r)$.
The alphabet simply reads $\{1-r,r\}$.
The $A(r)$ matrix in Eq.~\eqref{eq:amatrix} is lower triangular and can be solved iteratively.
The integral $I_1(r)$, which is the starting integral of our system of equations, is obtained by direct integration and we find
\begin{align}\label{eq:int1}
I_1(r) = e^{2 \epsilon \gamma_E}\,  \frac{r^{-\epsilon} (1-r)^{-3 \epsilon} \Gamma(1-\epsilon)}{2 \Gamma(1-3 \epsilon)}\, \theta(r)\theta(1-r)\, .
\end{align}
We notice that $I_2(r)$ and $I_3(r)$ satisfy the same differential equation. Hence, they will lead to identical results.
Starting from the result for $I_1(r)$, it is possible to compute $I_3(r)$ (and $I_2(r)=I_3(r)$) by solving the differential equation. The result is
\begin{align}\label{eq:int3}
    I_3(r) = r^{-2 \epsilon} e^{2 \epsilon  \gamma_E}\bigg( C_3(\epsilon) + \frac{r^\epsilon \,\Gamma(1-\epsilon) \, _2F_1(\epsilon, 3 \epsilon, 1+\epsilon,r)}{\Gamma(1-3 \epsilon)} \bigg)\, \theta(r)\theta(1-r)\, .
\end{align}
The $\epsilon$-dependent constant $C_3(\epsilon)$ can be fixed by requiring that the integral of $I_3(r)$ in the range $r\in [0,1]$ is equal to the parent integral where the $\delta(\omega- n_-k_1)$ constraint is removed. This latter integral is easily evaluated by direct integration. Specifically, we require that
\begin{align}
\int_0^1 dr \, I_3(r) = -\frac{e^{2 \epsilon \gamma_E}\,  2\,  \epsilon \, \Gamma(1-\epsilon)^2}{\Gamma(3-4 \epsilon)}\, ,
\end{align}
which fixes the $\epsilon$-dependent constant of $I_3(r)$ to be
\begin{align}
C_3(\epsilon) = -\frac{\Gamma(1+\epsilon)\Gamma(1-\epsilon)}{\Gamma(1-2 \epsilon)}\, .
\end{align}
We now focus on the integral $I_4(r)$ that satisfies a differential equation which involves both $I_1(r)$ and $I_3(r)$, as dictated by Eq.~\eqref{eq:diffeq}. Its solution reads
 \begin{align}\label{eq:int4b}
I_4(r) & = (1-r)^{-4 \epsilon} \, \bigg[ C_4(\epsilon) + 
\frac{e^{2 \epsilon \gamma_E}\epsilon\, \Gamma(1-\epsilon)}{\Gamma(1-3\epsilon)}
\int_1^r \, d r^\prime\, (1-r^\prime)^\epsilon (r^\prime)^{-1-\epsilon} & \nonumber \\
&\hspace{3.0cm}\times\, \bigg(1 - \frac{ 2  \epsilon (1 - r^\prime) 
\, _2F_1(1,1-2 \epsilon,2-3 \epsilon, 1-r^\prime)}{1-3 \epsilon}\bigg) \bigg]\, \theta(r)\theta(1-r)\, .
\end{align} 
The most complicated part of the integration concerns the hypergeometric function which we rewrite using its integral representation
\begin{align}
_ 2F_1(1, 1 - 2 \epsilon, 2 -3\epsilon, 1-r^\prime) & \equiv 
 \frac{\Gamma(2-3\epsilon)}{\Gamma(1 - 2 \epsilon) \, \Gamma(1-\epsilon)}\, 
 \int_{0}^{1} dt\, \frac{ t^{- 2 \epsilon} (1 - t)^{-\epsilon} } {1 - t \, (1-r^\prime)}\, .
\end{align}
Then, we make the variable transformation $r^\prime \to 1+ R \, (r-1)$ and integrate over the range $R\in [0,1]$. Finally, we carry out the integration over $t$ and arrive at the result
\begin{align}\label{eq:int4} \nn
I_4(r) =&  - (1-r)^{-4 \epsilon} \, e^{2 \epsilon \gamma_E }
\Gamma(1-\epsilon) \,
\bigg[\frac{2\,\Gamma(1-2\epsilon)\Gamma(1+\epsilon)}{\Gamma(1-4\epsilon)} \\
&\hspace{0.0cm} +\, \frac{\epsilon\, r^{-1-\epsilon}(1-r)^{1+\epsilon}}{(1+\epsilon)\Gamma(1-3\epsilon)}
\, _3 F_2\bigg(1,1-\epsilon, 1+\epsilon;\, 1-3\epsilon, 2 + \epsilon ;\frac{r-1}{r}\bigg) \bigg] \theta(r)\theta(1-r),
\end{align}
which retains exact $d$-dimensional dependence.
In the above equation the integration constant has already been fixed following the same procedure as used for the integral~$I_3(r)$.
We checked that the resulting expression for~$I_4(r)$ satisfies its initial differential equation.

Finally, we consider the last and most difficult integral of topology $\mathcal{B}$: $I_5(r)$.
The differential equation for $I_5(r)$ reads
\begin{align}
\frac{d I_5(r)}{d r} = \epsilon \bigg[ \frac{1}{r} I_1(r) + \frac{1}{r} I_3(r) + \frac{1}{r} I_4(r) -\frac{2}{r} I_5(r) \bigg]\, ,
\end{align}
and the solution has the following structure
\begin{align}\label{eq:i5diffeq}
I_5(r) = r^{-2 \epsilon}\bigg[  C_5(\epsilon)+ \int_1^r \, dr^\prime \, f_{I_5}(r^\prime, \epsilon) \bigg]\, \theta(r)\theta(1-r)\, .
\end{align}
We do not write explicitly the function $f_{I_5}(r^\prime, \epsilon)$ since it is too lengthy, but we know its exact expression in $d$-dimensions.
Unfortunately, we are not able to directly integrate $f_{I_5}(r^\prime, \epsilon)$ in $d$-dimensions since it involves a $_3F_2$ hypergeometric function. Nevertheless, we devise a technique which allows us to retain the dependence on $r^{- 2 \epsilon}$ terms, after $r^\prime$ integration, which is the relevant information that is needed to regularize the convolution integrals. Indeed these are the potential problematic contributions since the division by $r$ in the transformation to the non-canonical basis (see the last term of \eqref{eq:noncan}) will generate delta terms and plus distributions, after $\epsilon$ expansion, for the non-canonical integral $I^\prime_5(r)$.
Therefore, we need to treat these terms with care.
We follow the strategy of expanding the function $f_{I_5}(r^\prime, \epsilon)$ in the limit $r^\prime \to 0$ (up to finite order in $r^\prime$) and add and subtract this term in the following way
\begin{align}\label{eq:int5split}
\int_1^r \, dr^\prime \, f_{I_5}(r^\prime, \epsilon) &= \int_1^r \, dr^\prime \, \bigg(\lim_{r^\prime\to 0} f_{I_5}(r^\prime, \epsilon) \bigg) \, \nonumber \\
&+\int_1^r\, d r^\prime \, \bigg[ \lim_{\epsilon\to 0} \bigg( f_{I_5}(r^\prime, \epsilon) - \lim_{r^\prime\to 0} f_{I_5}(r^\prime, \epsilon)\bigg)\bigg] + \mathcal{O}(\epsilon^4)\, ,
\end{align}
where the $\epsilon$-limit in the above equation means that one needs to perform the $\epsilon\to 0 $ expansion up to the relevant order.
In Eq.~\eqref{eq:int5split} we split the integral in two terms, the first which we are able to integrate in $d$-dimensions exactly and the second which we need to $\epsilon$-expand before integration.
For the first term, we find
\begin{align}
\int_1^r \, dr^\prime \, \bigg(\lim_{r^\prime\to 0} f_{I_5}(r^\prime, \epsilon) \bigg) & = \, \frac{1}{2} e^{2 \epsilon \gamma_E} \frac{\Gamma(1+\epsilon) \Gamma(1-\epsilon)}{\epsilon} \bigg[ \frac{8 \epsilon^3 (r-1)}{\Gamma(2-2 \epsilon)} \, \nonumber \\
& + \frac{3 (r^\epsilon -1+ \epsilon^2 (3 (\epsilon-2)+r^\epsilon ((7 -3 \epsilon) r - 1)))}{(\epsilon^2-1) \Gamma(1-3 \epsilon) \Gamma(\epsilon)} \bigg].
\end{align}
The second term in \eqref{eq:int5split} contains terms that are non-singular in the $r\to 0$ limit and can be expressed in terms of standard \texttt{HPL}s \cite{Remiddi:1999ew} as follows
\begin{align}
& \int_1^r\, d r^\prime \, \bigg( \lim_{\epsilon\to 0} f_{I_5}(r^\prime, \epsilon) - \lim_{\epsilon \to 0}\lim_{r^\prime\to 0} f_{I_5}(r^\prime, \epsilon)\bigg)  =\, \nonumber \\
&\epsilon^2 \bigg( \frac{13}{12}\pi^2 + \frac{13}{2} (r-1) - \frac{13}{2} H(0,1;r)\bigg) + \epsilon^3 \bigg( \frac{25}{2} (1-r) +\frac{21}{2} \, r\, H(0;r) - \frac{17}{2} H(0,0,1;r) \, \nonumber \\
& - \frac{21}{2} H(0,1,0;r) -\frac{55}{2} H(0,1,1,r) + 15 \zeta_3\bigg) +\mathcal{O}(\epsilon^4)\, . 
\end{align}
We now have all the ingredients to construct the non-canonical integral $I^\prime_5(r)$ by using the last equation of \eqref{eq:noncan}  which combines the canonical integrals $I_4(r)$ and $I_5(r)$. The expression retains the exact $d$-dependence for terms of the type $r^{-1 - 2 \epsilon}$ and $(1-r)^{-1 - 4 \epsilon}$ which makes it suitable to be convoluted with $\mathcal{O}(\alpha_s)$ collinear functions avoiding
undefined convolutions at fixed-order accuracy.
In total we find
\begin{align}\label{eq:int5}
I^\prime_5(r) &= \Bigg[(1-r)^{-1-4\epsilon}\, \frac{2 e^{2\epsilon \gamma_E} \Gamma(1-2\epsilon) \Gamma (1-\epsilon) \Gamma(1+\epsilon)}{\epsilon^2 \Gamma (1-4\epsilon)}  \, \nonumber \\
& + r^{-1-2 \epsilon} \bigg(3\frac{e^{2 \epsilon \gamma_E} \big(1 +6\epsilon^2-3\epsilon^3  \big) \Gamma (1-\epsilon) \Gamma (1+\epsilon)}{2 (1-\epsilon) \epsilon^2 \Gamma(1-3 \epsilon) \Gamma ( 2 + \epsilon)}
\nonumber\\
&- \frac{ e^{2 \epsilon \gamma_E} \big(1-2 \epsilon +8 \epsilon^2\big) \Gamma (1-\epsilon) \Gamma(1+\epsilon)}{2 \epsilon^2 \Gamma (2-2 \epsilon)}\,+ \frac{13}{12} \big(\pi^2-6\big)  + \frac{30}{12} \epsilon \, \big (6\, \zeta_3+5\big ) \bigg)
\nonumber\\&
+\frac{1}{{6\, \epsilon (r-1) r}} \bigg(6 \epsilon (6 - 7 r) \text{Li}_2(r)  + 48\,  \epsilon (r-1) \ln^2(1-r) + r \big(3\, \epsilon \ln^2 r + \epsilon \pi^2  + 6 \ln r\big) \, \nonumber \\
& - 12 \ln(1-r) \big ( (r+1) \epsilon \ln r  + 2 (r-1) \big) \bigg)\Bigg] \, \theta(r)\theta(1-r)  +  \mathcal{O}(\epsilon)\, .
\end{align}
One can also expand the boundary singular terms in $\epsilon \to 0$ by using the relation
\begin{align}
x^{ - 1 - n\, \epsilon} = -\frac{\delta(x)}{(n \epsilon)} + \bigg[ \frac{1}{x} \bigg]_+  - (n \epsilon)  \bigg[ \frac{\ln x}{x} \bigg]_+ + \frac{(n \epsilon)^2}{2!} \bigg[ \frac{\ln^2x}{x} \bigg]_+ + \ldots\, ,
\end{align}
to find
\begin{align}\label{eq:int5nc}
I^\prime_5(r) = & \Bigg[-\frac{\delta(1-r)+ \delta(r)}{2 \epsilon^3} + \frac{1}{\epsilon^2}\bigg(2\bigg[\frac{1}{1-r}\bigg]_+ + \bigg[\frac{1}{r}\bigg]_+\bigg) + \frac{1}{12 \epsilon} \bigg( 5 \pi^2 \delta(1-r) -\pi^2 \delta(r) \, \nonumber \\
&- 96  \bigg[\frac{\ln (1-r)}{1-r}\bigg]_+ - 24 \bigg[\frac{\ln r}{r}\bigg]_+     -\frac{48 \ln (1-r)}{r} - \frac{12 \ln r}{1-r}\bigg)  \, \nonumber \\
&+ \frac{\zeta_3}{3} \big(28 \delta(1-r) - 5 \delta(r)\big )  -\frac{5  \pi^2}{3} \bigg[\frac{1}{1-r}\bigg]_+    +  \frac{\pi^2}{6} \bigg[\frac{1}{r}\bigg]_+  + 16 \bigg[ \frac{\ln^2(1-r)}{1-r}\bigg]_+  \, \nonumber \\
&+ 2 \bigg[\frac{\ln^2 r}{r}\bigg]_+ + 8 \frac{\ln^2(1-r)}{r} + \frac{2 (1+r)}{r(1-r)}\,  \ln(1-r) \ln(r) + \frac{\ln^2 r }{2 (r-1)}-\frac{7 \pi^2}{6} \, \nonumber \\
& + \frac{(6-7 r)}{(r-1) r} \bigg(\text{Li}_2(r)-\frac{\pi^2 r}{6}\bigg) \Bigg] \, \theta(r)\theta(1-r) + \mathcal{O}(\epsilon)\, .
\end{align}
In order to reproduce the cross section we need to integrate our results over $r$ in the range $r \in [0,1]$ and we obtain
\begin{align}
\int_0^1 \, dr \, I^\prime_5(r) = -\frac{1}{\epsilon^3} + \frac{7 \pi^2}{6 \epsilon} + \frac{62 \zeta_3}{3} + \, \ldots.
\end{align}
The integration constants are fixed similarly to $I_3$ and $I_4$. It turns out that they are zero up to the finite order in $\epsilon$.

We still need to discuss the calculation of the two MIs in Eq.~\eqref{eq:I6I7} which implement the constraint $\delta(\omega - n_-k_1 - n_-k_2)$ and the last MI in Eq.~\eqref{eq:I8} with the constraint $\delta(\omega_1-n_-k_1)\, \delta(\omega_2-n_-k_2)$. These integrals are carried out by direct integration in a straightforward way. For completeness we report the results below
\begin{align}
\hat{I}_6(\Omega,r) &= \Omega^2  \bigg(\frac{\mu}{\Omega}\bigg)^{4\epsilon} r^{1-2\epsilon}(1-r)^{ 1-2\epsilon} 
 \frac{e^{2\epsilon \gamma_E}\,\Gamma(1-\epsilon)^2}{\Gamma(2-2\epsilon)^2}
\, \theta(1-r)\,\theta(r),\, \\
\hat{I}_7(\Omega,r) &= \frac{1}{\Omega^2}\bigg(\frac{\mu}{\Omega} \bigg)^{4 \epsilon}\, r^{-1-2\epsilon} (1 -r)^{-1-2\epsilon}\, \theta(1-r)\,\theta(r)
\nonumber \\  & \hspace{0.75cm}\times
3 \, \frac{ e^{2\epsilon\gamma_E}}{\epsilon^2 }
\frac{\Gamma(1-\epsilon)}{\Gamma(1-3\epsilon) }
\,\, \, _3F_2\left( 
-\epsilon,-\epsilon,-\epsilon; -3\epsilon
,1-\epsilon ;1\right),\, \\
\hat{I}_8(\Omega,r_1, r_2) &=\bigg(\frac{\mu}{\Omega} \bigg)^{4 \epsilon} \frac{(1 - r_1 - r_2)^{1-2\epsilon}}{ r_1^{ \epsilon}
 \, r_2^{ \epsilon}}\,
  \frac{e^{2\epsilon\gamma_E}}{\Gamma(2-2\epsilon)}
 \,\,\theta(1 - r_1 - r_2) \theta(r_1)\theta(r_2) \,,
\end{align}
where $r_1=\omega_1/\Omega$ and $r_2=\omega_2/\Omega$.

\subsection{Results}
\label{sec:finalresults}
In this subsection we collect the final expressions for the bare soft functions which constitute the main results of our work. We retain the relevant $d$-dimensional dependence at the integration boundaries to avoid divergent convolutions when combining the soft functions with collinear functions \cite{Beneke:2019oqx} to fixed-order accuracy.
We refrain from expanding our results in $d \to 4$ since a consistent procedure for the renormalization of the soft functions beyond LL accuracy is not yet available in the literature.

Starting from the reduced result for $S^{(2)2r0v}_{1 }$ in Eq.~\eqref{eq:S1wk1reduced} expressed in terms of non-canonical MIs and using the transformations in Eqs.~\eqref{eq:intredef} and \eqref{eq:noncan}, it is then necessary to substitute the expressions for the canonical MIs in \eqref{eq:int1}, \eqref{eq:int3}, \eqref{eq:int4} and \eqref{eq:int5} to obtain
the explicit result for the real-real contribution to the $S_1$ soft function.
We do not report the complete expression for $S^{(2)2r0v}_{1 }(\Omega,\omega)$ here due to its length, but it is possible to easily reconstruct it from the information provided in the two subsections above.
With an analogous procedure we obtain the results for the $S_3$, $S_4$ and $S_5$ soft functions  
\begin{eqnarray}\label{eq:S3twoloop}
S^{(2) }_{3 }(\Omega,\omega) \!&=&\! 
 \frac{\alpha^2_s }{(4\pi)^2}
 C_FC_A
 \left( 
  \frac{\omega^{2}(\Omega-\omega)^{   2 }}{\mu^{4 }}
 \right)^{- \epsilon}\!\frac{2}{\omega}
 \frac{(1-\epsilon)}{  (3-2\epsilon)}
 \nonumber \\ && \times\frac{e^{2\epsilon \gamma_E}\,\Gamma(1-\epsilon)^2}{\Gamma(2-2\epsilon)^2}
 \theta(\Omega-\omega)\theta(\omega)\, ,\quad\quad
 \\
 \label{eq:S4complete}
 S^{(2)}_4(\Omega,\omega_1,\omega_2)
 \!&=&\!-\frac{\alpha^2_s}{(4\pi)^2}
C_FC_A\left(
\frac{ \omega_1^{  }
 \, \omega_2^{  }
 (\Omega-\omega_1  -\omega_2)^{ 2 }}{\mu^{4 }}\right)^{-\epsilon}
\frac{2      \omega_2  (1-\epsilon ) ( \omega_1 - \omega_2 )  }{( \omega_1 + \omega_2 )^4 }
\,\nonumber \\ && \times
 \frac{e^{2\epsilon\gamma_E}}{\Gamma(2-2\epsilon)}
 \,\,\theta(\Omega-\omega_1-\omega_2) \theta(\omega_1)\theta(\omega_2)\, ,
\\\label{S5result}
S^{(2) }_{5}(\Omega,\omega_1,\omega_2)\! &=& \! -\frac{\alpha^2_s}{(4\pi)^2}\left(C_F^2-\frac{1}{2}C_FC_A\right) 
\left(
\frac{ \omega_1^{  }
\, \omega_2^{  }
(\Omega-\omega_1  -\omega_2)^{ 2 }}{\mu^{4 }}\right)^{-\epsilon}
\frac{8 (1-\epsilon )   \,    \omega_2    }{( \omega_1 + \omega_2 )^3 }
\nonumber \\ && \times
\frac{e^{2\epsilon\gamma_E}}{\Gamma(2-2\epsilon)}
\,\,\theta(\Omega-\omega_1-\omega_2) \theta(\omega_1)\theta(\omega_2)\,.
\end{eqnarray}

In the above expressions the strong coupling constant is understood to be the renormalized coupling $\alpha_s\equiv \alpha_s(\mu)$ in the $\overline{\textsc{MS}}$ scheme obtained via the relation $Z_{\alpha} \alpha_s \mu^{2 \epsilon} = (4 \pi e^{-\gamma_E})^\epsilon \,  \alpha^0_s$, where $\alpha^0_s$ is the bare coupling constant and $Z_{\alpha} = 1 - \beta_0 \alpha_s/(4 \pi \epsilon)$ with $\beta_0 = \tfrac{11}{3}\, C_A - \tfrac{4}{6}\, n_f$. The discussion of the renormalization procedure for the soft functions is outside the scope of this paper due to the aforementioned divergent convolution problem.
However, one should keep in mind that, according to coupling renormalization, contributions proportional to $Z_\alpha S^{(1)}_1$ will appear at $\mathcal{O}(\alpha^2_s)$ and must be taken into account before operator renormalization.

\section{Comparison to fixed order results}
\label{sec:Validation}

\subsection{Next-to-next-to-leading order: soft contributions}
\label{sec:ValidationA}

After presenting the main results of this work in the section above, we now proceed with their validation through comparisons against cross-section level results which are available in the literature.
The soft functions calculated in this work carry a dependence on the convolution variables $\omega$ or $\omega_1,\omega_2$.
In order to evaluate their contributions to the cross section, one needs to carry out the convolution integrals of the $\omega$-dependent soft functions with their respective collinear functions.   

We begin with the contribution of $S_1$ by considering the relevant part of the factorization formula in \eqref{eq:3.24} with the variable transformation $r=\omega/\Omega$. We have 
\begin{eqnarray}\label{sig6s}
   \Delta^{dyn\,(2) 2r0v}_{{\rm{NLP-soft}}, S_{1 }}(z)=  4\, Q\,\Omega\,H^{(0)}(Q^2)\,\,
   \int dr  J^{(0)}_{1,1}\left(x_a(n_+p_A); r \, \Omega\right)
   S^{(2)2r0v}_{1 }\left(\Omega,r\right).
\end{eqnarray}
Inserting the result for the $S^{(2)2r0v}_{1 }\left(\Omega,r\right)$ soft function and the tree-level collinear function in~\eqref{eq:j011},  then integrating over the convolution variable~$r$ one finds to all orders in $\epsilon$:  
\footnote{We note that this result is accurate to all orders in $\epsilon$, whereas the part proportional to $I_{\rm{5}}(r)$ above is obtained only to finite order in the $\epsilon$ expansion. This is not surprising since here we are only interested in the final results after convolution, hence we can switch the order of integration and perform the convolution integral first.
We checked that the two results agree in the $\epsilon$ expansion.}
 \begin{eqnarray}\label{full}
\Delta^{dyn\,(2) 2r0v}_{{\rm{NLP-soft}}, S_{1}}(z)
&=& \, \frac{\alpha^2_s}{(4\pi)^2}
\left(\frac{\Omega^{4}}{\mu^{4}}\right)^{-\epsilon}
\Bigg( \,C^2_F  \, \frac{{32}}{\epsilon^3}\frac{e^{2\epsilon\gamma_E}
\Gamma(1-\epsilon)^2}{\Gamma(1-4\epsilon)}
 -{4}\,C_FC_A\,\frac{ e^{2 \epsilon \gamma_E  }  
 \Gamma (1-\epsilon )^2  }{\epsilon^3  (1-2 \epsilon )^2  \Gamma (1-4 \epsilon )}
 \nonumber\\ && \hspace{-2.0cm}\times
  \left(
\frac{  \left( 3 -25 \epsilon +50 \epsilon^2  -23 \epsilon^3 \right)}{(3-2 \epsilon  )}
  - \frac{ 3\Gamma (2-2 \epsilon )^2 }{\Gamma(1-\epsilon)\Gamma(1 -3\epsilon)}  \, _3 {F}_2(-\epsilon ,-\epsilon ,-\epsilon ;1-\epsilon ,-3 \epsilon ;1)\right)
  \nonumber\\ && 
  -{8}\,C_F n_f\frac{1}{\epsilon^2}
\frac{1}{(1-2\epsilon)^2(3-2\epsilon)}
\frac{e^{2\epsilon\gamma_E}\Gamma(2-\epsilon)^2}{\Gamma(1-4\epsilon)}\,\Bigg) .
 \end{eqnarray}
Setting the soft scale to $\Omega = Q (1-z)$, the scale $\mu = Q$, and finally expanding in $\epsilon$ we arrive at the following expression
\begin{eqnarray}\label{soft1CF2}
\Delta^{dyn\,(2) 2r0v}_{{\rm{NLP-soft}}, S_{1}}(z)&=&
\frac{ \alpha^2_s }{(4\pi)^2}\, \bigg\{ 
C^2_F \,\bigg[ \frac{32}{\epsilon^3}
-\frac{128 }{\epsilon^2}\ln (1-z)+\frac{1}{\epsilon}
\bigg( 256 \ln^2(1-z) -\frac{112 \pi^2}{3}\bigg)
\nonumber \\ && \hspace{1.5cm}
+\,\frac{32}{3} \Big(-32 \ln^3(1-z)+14 \pi^2 \ln (1-z) 
-62 \zeta_3 \Big)\bigg] \nonumber \\ &&
+ \,C_F C_A\, \bigg[\frac{8}{\epsilon^3}
-\frac{4 }{3 \epsilon^2}\Big(24 \ln (1-z )-11\Big)
-\frac{16}{9 \epsilon } \Big(-36 \ln^2(1-z )
\nonumber \\  && \hspace{1.5cm}
+33 \ln (1-z ) +6 \pi^2-16\Big) -\frac{256}{3} 
\ln^3(1-z )+\frac{352 }{3}\ln^2(1-z )
\nonumber \\  && \hspace{1.5cm}
+\,\frac{128}{3} \pi^2 \ln (1-z ) 
-\frac{1024}{9} \ln (1-z)-\frac{616 \zeta_3}{3}
-\frac{154 \pi^2}{9}+\frac{1484}{27}  \bigg]
\nonumber \\ &&
+C_F n_f\, \bigg[- \frac{8}{3\epsilon^2}
+\frac{32}{9\epsilon} \Big( 3\ln (1-z) -2 \Big) 
+\frac{4}{27} \Big(-144 \ln^2(1-z)
\nonumber \\ && \hspace{1.5cm} 
+\, 192 \ln(1-z) -122 +21\pi^2\Big) \bigg]
+\mathcal{O}\left(\epsilon\right)\bigg\}.
\end{eqnarray}
We recall that the complete contribution to $S_1$ proportional to $C_F C_A$ comprises an additional term, which stems from diagrams involving virtual-real corrections. Such contribution can be found in Eq.~(5.10) of \cite{Beneke:2019oqx}, and reads
\begin{eqnarray} \label{sig6s1r1vB}
\Delta^{dyn\,(2)1r1v}_{{\rm{NLP-soft}},S_{1,C_F C_A}}(z)&=& 
\frac{\alpha_s^2}{(4\pi)^2} 
\,C_FC_A \, \bigg[-\frac{8}{\epsilon^3}+
\frac{32 \ln (1-z)}{\epsilon^2}-
\frac{64 \ln^2(1-z)}{\epsilon }+\frac{28\pi^2}{3 \epsilon }
\nonumber \\
&&\hspace{-0cm}+\frac{256}{3} \ln^3(1-z)-\frac{112}{3} \pi^2 \ln (1-z)+\frac{448 \zeta_3}{3} +\mathcal{O}(\epsilon) \bigg]\,.
\end{eqnarray}
It is interesting to notice that the leading logarithmic contribution proportional to $C_F C_A$ cancels at cross section level when summing the double real, Eq.~\eqref{soft1CF2}, and virtual-real, Eq.~\eqref{sig6s1r1vB}, corrections. Such cancellation is expected, given that at order $n$ in $\alpha_s$ the leading logarithms in the cross-section are proportional to $C_F^n$ \cite{Beneke:2018gvs}.

After summing Eqs.~\eqref{soft1CF2} and \eqref{sig6s1r1vB} we obtain the complete contribution of $S_1$ to the partonic cross section at NNLO
\begin{eqnarray}\label{S1NNLOxs}
\Delta^{dyn\,(2)}_{{\rm{NLP-soft}}, S_{1}}(z) &=& 
\frac{\alpha^2_s}{(4\pi)^2} \bigg\{ C^2_F \,\bigg[ 
\frac{32}{\epsilon^3} -\frac{128}{\epsilon^2} \ln (1-z)
+\frac{1}{3\epsilon} \Big(768\ln^2(1-z) - 112 \pi^2 \Big)
\nonumber \\ && \hspace{2cm}
+\frac{32}{3} \Big(-32 \ln^3(1-z)+14 \pi^2 \ln (1-z) 
-62 \zeta_3 \Big)\bigg]  \nonumber \\ 
&&+\,C_F C_A \, \bigg[ \frac{44}{3 \epsilon^2}
-\frac{4}{9\epsilon}\Big( 132 \ln (1-z )- 64 + 3 \pi^2\Big)
+\frac{2}{27}\bigg(1584 \ln^2(1-z) \nonumber \\  && \hspace{1cm}
- 1536 \ln(1-z) + 72 \pi^2 \ln(1-z) 
+742 - 231 \pi^2-756\zeta_3 \bigg) \bigg] 
\nonumber \\
&&+\,C_Fn_f\, \bigg[- \frac{8}{3\epsilon^2}
+\frac{32}{9\epsilon} \Big( 3\ln (1-z) -2 \Big) 
+\frac{4}{27} \Big(-144 \ln^2(1-z) \nonumber \\ 
&& \hspace{2cm} + 192 \ln(1-z) - 122 +21 \pi^2 \Big)\bigg]
+\mathcal{O}\left(\epsilon\right) \bigg\}.
\end{eqnarray}

In addition to $S_1$, we need to take into account the contributions due to the 
other soft functions $S_3$, $S_4$, and $S_5$, which can be obtained integrating over the convolution variables the corresponding term in \eqref{eq:3.24}, similarly to what written in \eqref{sig6s}. $S_3$ and $S_4$ read 
\begin{eqnarray}\label{S3CFCAcrosssection}
\Delta^{dyn\,(2) 2r0v}_{{\rm{NLP-soft}}, S_{3}}(z)
= - \Delta^{dyn\,(2) 2r0v}_{{\rm{NLP-soft}}, S_{4}}(z)
&& \nonumber \\
&&\hspace{-5.0cm}
=\, {4}\, \frac{\alpha^2_s}{(4\pi)^2} \,C_FC_A\,
\left(\frac{\Omega^{4}}{\mu^{4}}\right)^{-\epsilon} \,
\frac{1}{\epsilon} \frac{(1-\epsilon)}{(1-2\epsilon)^2(3-2\epsilon)}
\frac{e^{2\epsilon\gamma_E}\Gamma(1-\epsilon)^2}{\Gamma(1-4\epsilon)}.\quad
\end{eqnarray}
Expanding in $\epsilon$ with $\Omega = Q (1-z)$ and $\mu=Q$ we obtain
\begin{eqnarray}\label{S3CFCAcrosssectionExp}
\Delta^{dyn\,(2) 2r0v}_{{\rm{NLP-soft}}, S_{3}}(z)
= -\Delta^{dyn\,(2) 2r0v}_{{\rm{NLP-soft}}, S_{4}}(z) && \nonumber \\
&&\hspace{-5.0cm}
=\, \frac{\alpha^2_s}{(4\pi)^2} \,C_FC_A\, 
\bigg[\frac{4}{3\epsilon}
-\frac{4}{9}\Big( 12\ln (1-z) -11 \Big)
+\mathcal{O}\left(\epsilon\right)\bigg].
\end{eqnarray} 
After convolution with the corresponding collinear function, $S_3$ and $S_4$ gives opposite contributions to the partonic cross section, such that they effectively cancel each other at this order.
The last contribution to the partonic cross section is given by the term involving $S_5$, and reads
\begin{eqnarray} \label{S5crosssection}
\Delta^{dyn\,(2) 2r0v}_{{\rm{NLP-soft}}, S_{5}}(z)
= {8}\, \frac{\alpha^2_s}{(4\pi)^2} \, 
\left( C_F^2 - \frac{1}{2} C_FC_A\right) \,
\left(\frac{\Omega^{4}}{\mu^{4}}\right)^{-\epsilon}
\frac{( 1 - \epsilon )}{\epsilon(1-2\epsilon)^2 } 
\frac{e^{2\epsilon \gamma_E}
\Gamma(1-\epsilon )^2}{\Gamma (1-4 \epsilon )}.
\end{eqnarray}
Setting $\Omega = Q (1-z)$, $\mu=Q$ and expanding in $\epsilon$ we find 
\begin{eqnarray} \label{S5crosssectionExp}
\Delta^{dyn\,(2) 2r0v}_{{\rm{NLP-soft}}, S_{5}}(z)
=  \, \frac{\alpha^2_s}{(4\pi)^2} \,
\left( C_F^2 - \frac{1}{2} C_FC_A\right) \,
\left[\frac{8}{\epsilon}- 32\ln(1-z)+24 
+\mathcal{O}(\epsilon) \right] \,.
\end{eqnarray}
We point out that the contributions to the cross section due to the soft functions starting at $\mathcal{O}(\alpha_s^2)$, namely $S_3$, $S_4$, and $S_5$, in~\eqref{S3CFCAcrosssectionExp} and~\eqref{S5crosssectionExp}, do not contain leading logarithmic terms. This confirms an assumption made in \cite{Beneke:2018gvs}, where it was claimed that a logarithmically enhanced off diagonal mixing of these soft functions with the single gluon soft function is not possible.

Results concerning the calculation of soft gluon contribution to the partonic Drell-Yan cross section within a diagrammatic approach have been presented in~\cite{Bonocore:2016awd}. The contribution due to double real soft radiation is provided in Eq.~(5.2) of~\cite{Bonocore:2016awd} and contains the contribution due to $\Delta^{dyn\,(2) 2r0v}_{{\rm{NLP-soft}}}$ presented here in Eq.~\eqref{soft1CF2}. However, a direct comparison is not straightforward, because the expression in Eq.~(5.2) of~\cite{Bonocore:2016awd} contains also NLP corrections due to the expansion of the phase space from the integration of the LP matrix element squared.   In the present approach, these are a part of the kinematic correction in Eq.~\eqref{kinPlusdyn} discussed in \cite{Beneke:2019oqx}. Similarly, the contribution due to virtual-real soft radiation given in Eq.~\eqref{sig6s1r1vB} of this paper is included in Eq.~(4.6) of~\cite{Bonocore:2016awd}. However, the two contributions cannot be compared directly, as Eq.~(4.6) of~\cite{Bonocore:2016awd} contains also the contribution due to hard and collinear loops. The problem can be overcome by comparing with the individual terms giving rise to Eq.~(4.6) and~(5.2) of \cite{Bonocore:2016awd}, provided by one of us, and we confirm that the whole contribution to the cross section due to the soft function $S_1$ in \eqref{S1NNLOxs} agrees with the cross-section level diagrammatic calculation of~\cite{Bonocore:2016awd}. Moreover, the contribution due to $S_5$ has not been considered in~\cite{Bonocore:2016awd}, and we validate its expression against an in-house calculation performed with the expansion-by-regions method \cite{Beneke:1997zp}.

\subsection{Next-to-next-to-leading order: complete contribution}
The result obtained in this paper, together with the results given in \cite{Beneke:2019oqx}, can be compared with the reference \cite{Hamberg:1990np} which gives the full NNLO contribution to the Drell-Yan process. To this end, we recall that within the present approach the full partonic cross section is given according to Eq.~\eqref{kinPlusdyn}, that is, as the sum of a dynamic and a kinematic contribution. Writing explicitly all the terms contributing to the cross section, we have\footnote{We recall that we drop the indices $q\bar q$, namely, $\Delta_{\rm NLP}^{(2)} \equiv \Delta_{q\bar q\, \rm NLP}^{(2)}$.} 
\be\label{fullNNLO}
\Delta_{\rm NLP}^{(2)} = \Delta^{kin\,(2)}_{{\rm{NLP}}}(z) 
+ \Delta^{dyn\,(2)}_{{\rm{NLP-coll}}}(z)
+ \Delta^{dyn\,(2)}_{{\rm{NLP-hard}}}(z)
+\Delta^{dyn\,(2)}_{{\rm{NLP-soft}}}(z).
\ee
The first three terms have been calculated in \cite{Beneke:2019oqx}, and are reported explicitly in Eqs.~\eqref{eq:NNLOkin}, \eqref{sig9} and \eqref{sig13} of App.~\ref{appNNLOterms}.
The last term is given by the sum of the $S_1$, $S_3$, $S_4$ and $S_5$ contributions given in \eqref{S1NNLOxs}, \eqref{S3CFCAcrosssectionExp} and \eqref{S5crosssectionExp} respectively. Explicitly, it reads
\begin{eqnarray}\label{SNNLOxs}
\Delta^{dyn\,(2)}_{{\rm{NLP-soft}}}(z) &=& 
\frac{\alpha^2_s}{(4\pi)^2} \bigg\{ C^2_F \,\bigg[ 
\frac{32}{\epsilon^3} -\frac{128}{\epsilon^2} \ln (1-z)
+\frac{1}{3\epsilon} \Big(768\ln^2(1-z) +24 - 112 \pi^2 \Big)
\nonumber \\ && \hspace{0cm}
+\frac{8}{3} \Big(-128 \ln^3(1-z)-12 \ln(1-z) 
+56 \pi^2 \ln (1-z) +9 -248 \zeta_3 \Big)\bigg]  \nonumber \\ 
&&\hspace{-1.0cm}+\,C_F C_A \, \bigg[ \frac{44}{3 \epsilon^2}
-\frac{4}{9\epsilon}\Big(132 \ln (1-z )- 55 + 3 \pi^2\Big)
+\frac{2}{27}\Big( 1584 \ln^2(1-z) \nonumber \\  && \hspace{1cm}
- 1320\ln(1-z) + 72 \pi^2\ln(1-z) 
+ 580 - 231 \pi^2 - 756\zeta_3 \Big)\bigg] 
\nonumber \\
&&\hspace{-1.0cm}
+\,C_Fn_f\, \bigg[- \frac{8}{3\epsilon^2}
+\frac{32}{9\epsilon} \Big( 3\ln (1-z) -2 \Big) 
+\frac{4}{27} \Big(-144 \ln^2(1-z) \nonumber \\ 
&& \hspace{2cm} + 192 \ln(1-z) - 122 +21 \pi^2 \Big)\bigg]
+\mathcal{O}\left(\epsilon\right) \bigg\}.
\end{eqnarray}
Substituting this result along with expressions in Eqs.~\eqref{eq:NNLOkin},~\eqref{sig9} and~\eqref{sig13}  into \eqref{fullNNLO}, we arrive at the full NLP NNLO correction for the $q\bar q$ partonic channel of the Drell-Yan process at threshold
\begin{eqnarray}\label{bareNNLOxs}
\Delta^{(2)}_{{\rm{NLP}}}(z) &=& 
\frac{\alpha^2_s}{(4\pi)^2} \bigg\{ C^2_F \,\bigg[ 
 -\frac{16}{\epsilon^2} \Big(4 \ln (1-z) + 1\Big)
+\frac{1}{3\epsilon} \Big(576\ln^2(1-z) -336 \ln(1-z)
\nonumber \\ && \hspace{0cm}
-564 - 32 \pi^2 \Big) 
+\frac{4}{3} \Big(-224 \ln^3(1-z) + 306 \ln^2(1-z)
+285 \ln(1-z) \nonumber \\ && \hspace{0cm}
+72 \pi^2 \ln(1-z) -288 -14 \pi^2 
-384 \zeta_3 \Big)\bigg] \nonumber \\
&&\hspace{-1.0cm}+\,C_F C_A \, \bigg[ \frac{44}{3 \epsilon^2}
-\frac{4}{9\epsilon}\Big( 132 \ln (1-z ) - 166 + 3 \pi^2\Big)
+\frac{2}{27}\Big( 1584 \ln^2(1-z) \nonumber \\  && \hspace{1cm}
- 3714\ln(1-z) + 72 \pi^2\ln(1-z) 
+ 1402 - 267 \pi^2 - 756\zeta_3 \Big)\bigg] 
\nonumber \\
&&\hspace{-1.0cm}
+\,C_Fn_f\, \bigg[- \frac{8}{3\epsilon^2}
+\frac{8}{9\epsilon} \Big( 12\ln (1-z) -14 \Big) 
+\frac{4}{27} \Big(-144 \ln^2(1-z) \nonumber \\ 
&& \hspace{2cm} + 336 \ln(1-z) - 164 +21 \pi^2 \Big)\bigg]
+\mathcal{O}\left(\epsilon\right) \bigg\}.
\end{eqnarray}
This is the non-singlet contribution to the $q\bar q$ unrenormalized partonic cross section, expressed in terms of the bare coupling constant. To compare with \cite{Hamberg:1990np} we need to write it in terms of the UV-renormalized coupling constant and remove the initial state collinear singularities at cross-section level. These procedures amount to adding the following counterterms to the full NNLO cross section:
\bea\label{xsRenorm} \nonumber
\Delta^{(2)}_{{\rm ren}}(z)
 &=& \Delta^{(2)}(z) + 
 \bigg(\frac{\alpha_s}{4\pi}\bigg)^2 
 \bigg[- \frac{1}{2\epsilon^2} 
\bigg( P_{qq}^{0} \otimes P_{qq}^{0} 
- P_{qq}^{0} \beta_0 \bigg)
+ \frac{1}{2\epsilon} \bigg(P_{qq}^{1,\rm NS} \\
&&\hspace{-2.0cm}+\, 2 P_{qq}^{0} \otimes \Delta^{(1)}(z)|_{\epsilon^0}
- 2 \Delta^{(1)}(z)|_{\epsilon^0} \beta_0 \bigg) 
- 2 P_{qq}^{0} \otimes \Delta^{(1)}(z)|_{\epsilon}
+ 2 \Delta^{(1)}(z)|_{\epsilon} \beta_0 \bigg],
\eea 
when both sides are evaluated in terms of the renormalized coupling constant, for $\mu_f = \mu_r = Q$. In this equation the symbol $\otimes$ indicates convolution\footnote{We use the program MT \cite{Hoschele:2013pvt} to evaluate convolutions of plus distributions.}:
\be
(f_1 \otimes f_2 )(x) = \int_0^1 dx_1 \, dx_2 
\, \delta(x - x_1 \, x_2) f_1(x_1) f_2(x_2),
\ee
furthermore, by $\Delta^{(1)}(z)|_{\epsilon^0}$
and $\Delta^{(1)}(z)|_{\epsilon}$ we indicate respectively the coefficient of the $\epsilon^0$ and $\epsilon$ terms of the NLO $q\bar q$ Drell-Yan correction, and $P_{qq}^0$ and $P_{qq}^{1,\rm NS}$ represent the one- and two-loop Altarelli-Parisi splitting functions, that we provide for completeness in App. \ref{AppAP}. We note that Eq.~\eqref{xsRenorm} is valid to all powers in $(1-z)$. Expanding all terms 
in $(1-z)$ and selecting the NLP contribution we finally get the finite partonic cross section 
\begin{eqnarray}\label{bareNNLOxs2}
\Delta^{(2)}_{{\rm{NLP},\, \rm{ren}}}(z) &=& 
\frac{\alpha^2_s}{(4\pi)^2} \bigg\{ C^2_F \,\bigg[ \,
\frac{4}{3} \Big(-96 \ln^3(1-z) +186 \ln^2(1-z)
+213 \ln(1-z) \nonumber \\ && \hspace{3cm}
+16 \pi^2 \ln(1-z) -96 -16 \pi^2 
-192 \zeta_3 \Big)\bigg] \nonumber \\
&&\hspace{0.0cm}+\,C_F C_A \, \bigg[ \,
\frac{4}{27}\Big( 396 \ln^2(1-z) 
- 1461\ln(1-z) + 36 \pi^2\ln(1-z) 
\nonumber \\ && \hspace{3cm}
+ 701 - 84 \pi^2 - 378\zeta_3 \Big)\bigg] 
\\ \nonumber
&&+\,C_F n_f\, \bigg[ \,
\frac{8}{27} \Big(-36 \ln^2(1-z) 
+ 132 \ln(1-z) - 82 +6 \pi^2 \Big)\bigg]
+\mathcal{O}\left(\epsilon\right) \bigg\},
\end{eqnarray}
which agrees with the NLP content of Eq.~(B.7) in \cite{Hamberg:1990np}, for $\mu_f = \mu_r = Q$, 
namely $\Delta_{q\bar q}^{(2),C_A} + \Delta_{q\bar q}^{(2),C_F} + \Delta_{q\bar q,A^2}^{(2)} + 2 \Delta_{q\bar q,AC}^{(2)}$ expanded to NLP. These terms are provided explicitly in \cite{Hamberg:1990np} in Eqs. (B.30) -- (B.33).

\subsection{Next-to-next-to-next-to-leading order}
Recently, partial  results for the NLP expansion of the $C_F^3$ contribution to the N$^3$LO Drell-Yan cross section were calculated in \cite{Bahjat-Abbas:2018hpv} using the expansion-by-region method. We are able to compare to this result by combining the result of the collinear function $J^{(1)}_{1,1}$ at $\mathcal{O}(\alpha_s)$ \cite{Beneke:2019oqx} and the calculation of the $\mathcal{O}(\alpha^2_s)$ soft function $S_1^{(2)}$ obtained in this paper.
To this end, we now focus on the following part of the factorization formula expanded to the third order in the coupling constant
\begin{eqnarray}\label{eq:3.43bN3LO}
\Delta^{dyn\,(3)}_{{\rm{NLP-coll}}, \,C_F^3}(z)&=& 4 Q \int d\omega
\,{J}^{(1)}_{1,1}(x_a\,n_+p_A; \omega)  
\,{S}^{(2)}_{1, C_F^2}(\Omega; \omega) \,.
\end{eqnarray}
For the one-loop collinear function we use the $C_F$ part of the result given in~\eqref{J1} after the colour generator and Dirac-index Kronecker-symbol are removed. The relevant two-loop soft function piece reads
\begin{eqnarray}\label{4.1.2.11}& &S^{(2)2r0v}_{1,C_F^2 }(\Omega,\omega) =8 \frac{\alpha^2_s}{(4\pi)^2} \,C^2_F\,\left(\frac{\omega\,(\Omega-\omega)^{3}}{\mu^{4 }}\right)^{-\epsilon}\frac{1}{\omega}\, \frac{1}{\epsilon^2}\frac{e^{2\epsilon\gamma_E}\Gamma(1-\epsilon)}{\Gamma(1-3\epsilon)}\theta(\Omega-\omega) \theta(\omega).\quad\,\,
\end{eqnarray}
We perform the convolution according to~\eqref{eq:3.43bN3LO} and arrive at the following $d$-dimensional result
\begin{eqnarray}\label{CFNNNLO}
\Delta^{dyn\,(3)}_{{\rm{NLP-coll}},\,C_F^3}(z)&=& 32\frac{  \alpha_s^3 }{(4\pi)^3} 
 C_F^3 
  \left(\frac{Q\,\Omega^5 }
{\mu^6}\right)^{-\epsilon }
\,
\frac{1}{\epsilon^4} (-4 +   7\epsilon+\epsilon^2)
 \nonumber\\  &&
\times 
\frac{  e^{3 \epsilon\gamma_E  }   \Gamma(1+\epsilon) \Gamma(1-\epsilon)^2 \Gamma(1-2 \epsilon)}{\Gamma(1-5 \epsilon) \Gamma(3-2 \epsilon)}.
\end{eqnarray}
Setting   $\Omega=Q(1-z)$ and $\mu=Q$, and expanding in $\epsilon$ we find
\begin{eqnarray}\label{N3Loexp}
\Delta^{dyn\,(3)}_{{\rm{NLP-coll}},\,C_F^3}(z)&=&  \frac{  \alpha_s^3 }{(4\pi)^3} 
 C_F^3\bigg[ 
-\frac{64}{\epsilon^4}+\frac{80 (4 \ln (1-z)-1)}{\epsilon^3}
+\frac{16}{\epsilon^2} \bigg(-50 \ln^2(1-z)\nonumber\\  && +25 \ln (1-z)+7 \pi^2-6\bigg)
+\frac{1}{\epsilon}\bigg(\frac{4000}{3} \ln^3(1-z)
 -1000 \ln^2(1-z)\nonumber\\  &&
 -560 \pi^2 \ln (1-z)+480 \ln (1-z)+2624 \zeta_3+140 \pi^2-128 \bigg)
\nonumber\\  && 
 -\frac{5000}{3}  \ln^4(1-z)+\frac{5000}{3} \ln^3(1-z)+1400 \pi^2 \ln^2(1-z)
 \nonumber\\  && 
 -1200 \ln^2(1-z)-700 \pi^2 \ln (1-z)+640 \ln (1-z)
 \nonumber\\  && 
 +\zeta_3 (3280-13120 \ln (1-z))
 +\frac{62 \pi^4}{5}+168 \pi^2-192 \bigg].
\end{eqnarray}
Our expanded result in the equation above agrees with Eq.~(45) of~\cite{Bahjat-Abbas:2018hpv} up to the finite constant terms which are not reported there and a factor of two accounting for the anticollinear contribution.

\subsection{Cusp anomalous dimension}

In addition to the checks performed at the  cross-section level in the two sections above, we use the leading pole of the two-loop soft function $S_1$ to extract the first diagonal entry of the anomalous dimension matrix defined in Eq.~(3.50) of \cite{Beneke:2018gvs} finding agreement.
Currently, the resummation beyond LL is hampered by the appearance of endpoint divergent convolutions \cite{Beneke:2019oqx}. However, once cured, the results obtained in this work will be useful to extract the soft anomalous dimension matrix beyond LL accuracy.

\section{Conclusions}
\label{sec:summary}

In this article, we calculated the real-real contributions to the NLP generalized soft functions, which enter the bare factorization theorem for the Drell-Yan process in the threshold region \cite{Beneke:2019oqx}.
This allowed us to complete the comparison of the NNLO Drell-Yan cross-section up to NLP against existing fixed-order results.

The generalized soft functions, listed in Eqs.~\eqref{eq:3.23} -- \eqref{eq:3.27}, contain a dependence on additional convolution variables $\omega$ or $\omega_1, \omega_2$ with respect to the LP soft function. We carried out the calculation by employing methods developed for fixed-order calculations such as the reduction to MIs and the use of the differential equations for the direct evaluation of the MIs.
Our results retain the exact $d$-dimensional dependence on the convolution variables at the integration boundaries which allows us to perform the convolution integrals with collinear functions at fixed-order accuracy. Given the current issues stemming from the expansion in $d\to 4$ of the soft and collinear functions before the convolution is performed, we leave the non-trivial study of the renormalization procedure of the generalized soft functions for future work.

We showed that combining the soft functions with their respective collinear functions, as prescribed by the NLP factorization theorem, and performing the $d$-dimensional integrals yields the correct NNLO cross-section expressions up to NLP in the threshold expansion.
In addition, we reproduced partial N$^3$LO results available in the literature.
We also confirmed the result for the diagonal entry of the anomalous dimension computed in \cite{Beneke:2018gvs}, and explicitly validated to NNLO the assumption made in \cite{Beneke:2018gvs} that the soft functions beginning at $\mathcal{O}(\alpha_s^2)$ do not contribute to the LL series.
This is the first time that NLP soft functions are calculated to $\mathcal{O}(\alpha^2_s)$ and we hope that further investigations of their intricate structure can shed light on the endpoint divergent convolution problem currently prohibiting the NLP resummation of the threshold logarithms beyond LL accuracy for the Drell-Yan process.

\subsubsection*{Acknowledgments} 
 
We would like to extend our warmest thanks to Martin Beneke, Robert Szafron, and Chris Wever for discussions.
The work of A.B. is supported by the ERC Starting Grant REINVENT-714788. S.J. is supported by the UK Science and Technology Facilities Council (STFC) under grant ST/T001011/1 and received funding from the Bundesministerium f\"ur  Bildung und Forschung (BMBF) grant no. 05H18WOCA1. 
L.V. is supported 
by Fellini - Fellowship for Innovation at INFN, funded by the 
European Union's Horizon 2020 research programme under the 
Marie Sk\l{}odowska-Curie Cofund Action, grant agreement no. 754496. Figures were drawn with \texttt{Jaxodraw}~\cite{Binosi:2008ig}. Calculations were done in part with \texttt{FORM}~\cite{Vermaseren:2000nd} and expansions with \texttt{HypExp} \cite{Huber:2005yg}.

\begin{appendix}

\section{Collinear functions}
\label{app:collinearfns}
For completeness, we provide the results for the collinear functions, $J_i$ in \eqref{eq:3.24}, obtained in \cite{Beneke:2019oqx}. 
The tree-level 
collinear functions corresponding to insertions of power suppressed Lagrangian terms at a single position are given by 
\begin{eqnarray}
{J}^{(0)}_{1;\gamma\beta}(n_+q,n_+p;\omega)
&=&     \delta_{\beta\gamma } 
\left(-\frac{1}{n_+p} \delta(n_+q -n_+p)  
+2 \,\frac{\partial}{\partial n_+q}
\delta(n_+q -n_+p) \right),\qquad\,\,
\label{eq:J1fn}
\\ 
{J}^{\mu\nu,(0)}_{2;\gamma\beta}(n_+q,n_+p;\omega) &=&
-\frac{1}{2} \frac{1}{n_+p}  \,
\big[\gamma^{\mu}_{\perp}\gamma^{\nu}_{\perp} \big]_{\gamma\beta}
\,\delta(n_+q -n_+p)\,,
\quad
\\ 
{J}^{(0)}_{3;\gamma\beta}(n_+q,n_+p;\omega)
&=&     \delta_{\beta\gamma } 
\left(-\frac{1}{n_+p} \delta(n_+q -n_+p)  
+2 \,\frac{\partial}{\partial n_+q}
\delta(n_+q -n_+p) \right).\qquad\,\,
\label{eq:J3fn}
\end{eqnarray}
It is useful to write the $J_1(n_+p,x_a\,n_+p_A; \omega )$
collinear function in terms of two scalar components in the following way
\begin{eqnarray}\label{eq:collFuncDecomp}
{J}^{}_{1;\gamma\beta}
\left(n_+p,x_a\,n_+p_A; \omega \right) 
&=& \delta_{\gamma\beta}\,
\, \bigg[ {J}_{1,1}
\left(x_an_+p_A; \omega \right) 
\delta(n_+p-x_a n_+p_A)
\nn \\ && \hspace{-2cm} + \, {J}_{1,2}
\left(x_an_+p_A; \omega \right) 
\frac{\partial}{\partial (n_+p)} 
\delta(n_+p-x_a n_+p_A) \bigg]\,.
\end{eqnarray}
such that
\bea
{J}^{(0)}_{1,1}\left(n_+p;\omega\right) &=& 
-\frac{1}{n_+p}\,,\label{eq:j011}\\[1ex]
{J}^{(0)}_{1,2}\left(n_+p;\omega \right) &=& 2\,.
\label{eq:j012}   
\eea
Next, we write the tree-level collinear functions with two $\omega_i$ variables, corresponding to two time-ordered product insertions of the $\mathcal{O}(\lambda)$
Lagrangian terms. We have 
\begin{eqnarray}
J^{\mu\nu,AB\,(0)}_{4;\gamma\beta  ,fb }
\left(n_+q,n_+p;\omega_1,\omega_2\right) &=&\frac{2g^{\mu\nu}_{\perp}}{n_+p\,(\omega_1+\omega_2   )^2}
\left(  \, \omega_1   
\, \mathbf{T}^{A}\mathbf{T}^{B}  +\omega_2 \, \mathbf{T}^{B}_{ }\mathbf{T}^{A} 
\, \right)_{fb}\nonumber\\
&&\times\,  \,\delta(n_+q -n_+p)\,
\\&\equiv&\label{eq:J4tree} J^{\mu\nu,AB\,(0)}_{4; fb }
\left(n_+p;\omega_1,\omega_2\right)\delta_{\gamma\beta}\delta(n_+q -n_+p),
\end{eqnarray}
and
\begin{eqnarray}
\label{eq:twoquarkcolfunc}
{J}^{f k_1 k_2 e\,(0)}_{5;\gamma\sigma\lambda\beta }
(n_+q,n_+p; \omega_1, \omega_2)&=&
-\mathbf{T}^A_{f k_2}\mathbf{T}^A_{k_1 e}\,
\frac{1}{n_+p}
\frac{  \omega_2}{(\omega_1+\omega_2)}
\frac{\slashed{n}_{-\gamma\eta}}{2}
\gamma^{\mu}_{\perp,\eta\sigma}
\gamma^{}_{\perp\mu,\lambda\beta} \,\delta(n_+q -n_+p)
\nonumber \\ && \hspace{-2cm}
+\,2 \,\mathbf{T}^{K}_{fe} \textbf{T}^{K}_{k_1k_2}
		\frac{ \omega_1 \omega_2}{(\omega_1+\omega_2)^2}    
	\slashed{n}_{-\lambda\sigma}
\delta_{\gamma\beta}  \,\frac{\partial}{\partial n_+q}
\delta(n_+q -n_+p)\,.\qquad
\end{eqnarray}
We also give the one-loop collinear function corresponding to the only soft function starting at $\mathcal{O}(\alpha_s)$. The result  is
\begin{eqnarray}
\label{J1}
&& {J}^{(1)}_{1;\gamma\beta}
\left(n_+q, n_+p; \,\omega \right) = 
\frac{\alpha_s}{4 \pi} 
\delta_{\gamma\beta}  
\, \frac{1}{(n_+p) } \left(\frac{n_+p\,\omega}
{\mu^2}\right)^{-\epsilon }
\frac{e^{\epsilon\,\gamma_E}\,\Gamma[1+\epsilon ]
\Gamma [1-\epsilon]^2}
{(-1+\epsilon)(1+\epsilon) \Gamma [2-2 \epsilon ]}
\nonumber\\ 
&& \hspace{1.5cm}\times  
\left( C_F\left(-\frac{4}{\epsilon}+3 
+8\epsilon+\epsilon^2  \right)
-  C_A \left(-5+{8}{\epsilon}
+\epsilon^2 \right)\right) \delta(n_+q -n_+p).\qquad
\end{eqnarray}

\section{Matrix elements}
\label{sec:matrixelements}
To generate our starting cross-section level expressions we require the following results for the power suppressed matrix elements
 \begin{eqnarray}\label{14.29}
\langle g^{K_1}(k_1)g^{K_2}(k_2) |\mathbf{T}\left[ 
 Y^{\dagger}_-(0)Y_+(0) \, \frac{i\partial_{\perp}^{\mu}}{in_-\partial}
\mathcal{B}^+_{\mu_\perp}(z_-) \,
\right] 
 |0\rangle &=& \nonumber\\ && \hspace{-7cm }
g_s^2 \mathbf{T}^{K_2}
\,\mathbf{T}^{K_1}\,
\frac{1}{(n_-k_1)}\frac{n_-^{\eta_2}}{(n_-k_2)}  \,\left[ k^{\eta_1}_{1\perp}
-\frac{k^2_{1\perp}}{(n_-k_1)}n_-^{\eta_1} \right] 
\epsilon^*_{\eta_1\,}(k_1)\epsilon^*_{\eta_2\,}(k_2)
\, e^{iz_-k_1}
\nonumber\\ && \hspace{-7.3cm }
+g_s^2 \mathbf{T}^{K_1}
\,\mathbf{T}^{K_2}\,
\frac{1}{(n_-k_2)} \frac{n_-^{\eta_1}}{(n_-k_1)} \,\left[ k^{\eta_2}_{2\perp}
-\frac{k^2_{2\perp}}{(n_-k_2)}n_-^{\eta_2} \right] 
\epsilon^*_{\eta_1\,}(k_1)\epsilon^*_{\eta_2\,}(k_2)
\, e^{iz_-k_2}
 \nonumber\\ && \hspace{-7.3cm }
-g_s^2 \mathbf{T}^{K_2}
\,\mathbf{T}^{K_1}\,
\frac{1}{(n_-k_1)}\frac{n_+^{\eta_2}}{(n_+k_2)}  \,\left[ k^{\eta_1}_{1\perp}
-\frac{k^2_{1\perp}}{(n_-k_1)}n_-^{\eta_1} \right] 
\epsilon^*_{\eta_1\,}(k_1)\epsilon^*_{\eta_2\,}(k_2)
\, e^{iz_-k_1}
\nonumber\\ && \hspace{-7.3cm }
-g_s^2 \mathbf{T}^{K_1}\,
\mathbf{T}^{K_2}\,
\frac{1}{(n_-k_2)} \frac{n_+^{\eta_1}}{(n_+k_1)} \,\left[ k^{\eta_2}_{2\perp}
-\frac{k^2_{2\perp}}{(n_-k_2)}n_-^{\eta_2} \right] 
\epsilon^*_{\eta_1\,}(k_1)\epsilon^*_{\eta_2\,}(k_2)
\, e^{iz_-k_2}
\nonumber\\&& \hspace{-7.3cm}+
 g_s^2 \,if^{K_1K_2K}\mathbf{T}^{K} 
  \frac{1}{n_-(k_1+k_2)} 
 \Bigg(
  -\frac{      \,\left(k^{\eta_2}_{1\perp}+k^{\eta_2}_{2\perp}\right) n_-^{\eta_1}}{(n_-k_1) }
  + \frac{ \left(k^{\eta_1}_{1\perp}+k^{\eta_1}_{2\perp}\right)\,n_-^{\eta_2}}{(n_-k_2 )}
\nonumber \\ 
 &&\hspace{-6.7cm} 
 -
 \frac{ n_-^{\eta_1}n_-^{\eta_2}}{ n_-(k_1+  k_2)  (n_-k_1)(n_-k_2)}
 \Big[  (n_-k_1)\,\left(k^{2}_{1\perp}+k_{1\perp}\cdot k_{2\perp}\right)
 \nonumber \\ 
 &&\hspace{-5cm}    -(n_- k_2 )  \left(k_{2\perp}\cdot k^{}_{1\perp}+k^{2}_{2\perp}\right)  \Big]
\Bigg)
    \epsilon^{*}_{\eta_1}(k_1)
 \epsilon^{*}_{\eta_2}(k_2)\,e^{iz_-(k_1+k_2)}
 \nonumber \\ &&\hspace{-7.3cm}
+g_s^2if^{K_1K_2K}\mathbf{T}^{K}\, 
\frac{1}{(n_-(k_1+k_2))^2 }  \frac{1}{(k_1+k_2)^2}
\, \Bigg(\Big[{n}_{- }^{\eta_1} (2k_1+k_2)^{\eta_2} 
\nonumber \\ &&\hspace{-5cm}
-{n}_{-   }^{\eta_2} (k_1+2k_2)^{\eta_1}
-g^{\eta_2  \eta_1 } 
({n}_{-}(k_1-k_2))\Big](k_{1\perp}+k_{2\perp})^2 
\nonumber\\&&
\hspace{-6cm}
+
\Big[(k_{1\perp}^{\eta_1}+k_{2\perp}^{\eta_1}) (-2k_1-k_2)^{\eta_2}
+(k_{1\perp}^{\eta_2}+k_{2\perp}^{\eta_2}) (k_1+2k_2)^{\eta_1} \nonumber \\ &&\hspace{-7cm}
+g^{\eta_2  \eta_1 } \big(k^2_{1\perp }-k^2_{2\perp }\big)
 \Big]({n}_{-}(k_1+k_2)) \Bigg) 
    \epsilon^{*}_{\eta_1}(k_1)
 \epsilon^{*}_{\eta_2}(k_2)\,e^{iz_-(k_1+k_2)},
\end{eqnarray}
\begin{eqnarray}\label{14.29quarks}
\langle q (k_1)\bar{q}(k_2) |\mathbf{T}\left[ 
 Y^{\dagger}_-(0)Y_+(0) \, \frac{i\partial_{\perp}^{\mu}}{in_-\partial}
\mathcal{B}^+_{\mu_\perp}(z_-) \,
\right] 
 |0\rangle &=&
g_s^2
 \frac{1}{(n_-(k_1+k_2))^2}  \,\mathbf{T}^{B}_{}\,
\nonumber\\&&\hspace{-4cm }\times  
\left(  {n}_{-}(k_1+k_2) (k_{1\perp\nu}+k_{2\perp\nu})
-   (k_{1\perp}+k_{2\perp})^{2} {n}_{-\nu}  \right)  
\nonumber\\&&\hspace{-4cm }\times \frac{1}{(k_1+k_2)^2}   
\bar{u}(k_1) \mathbf{T}^{B}_{}
\gamma^{\nu} v(k_2) \,  e^{iz_-(k_1+k_2)},
\end{eqnarray}
and 
 \begin{eqnarray}\label{14.29ghosts}
\langle {c}^{K_1} (k_1) \bar{{c}}^{K_2}(k_2) |\mathbf{T}\left[ 
 Y^{\dagger}_-(0)Y_+(0) \, \frac{i\partial_{\perp}^{\mu}}{in_-\partial}
\mathcal{B}^+_{\mu_\perp}(z_-) \,
\right] 
 |0\rangle &=& 
g_s^2
 \frac{1}{(n_-(k_1+k_2))^2}  \, if^{K_1BK_2}  \mathbf{T}^{B}_{}\,
\nonumber\\&& \hspace{-4cm }\times  
\left( {n}_{-}(k_1+k_2) (k_{1\perp\nu}+k_{2\perp\nu}) -  (k_{1\perp}+k_{2\perp})^{2} {n}_{-\nu}   \right)  
\nonumber\\&& \hspace{-4cm }\times \frac{1}{(k_1+k_2)^2}\,   
k_1^{\nu}   \,
 e^{iz_-(k_1+k_2)}\,.
\end{eqnarray}
We also require the LP amplitudes
\begin{eqnarray}\label{eq:LPtwogluonab}
\langle 0| 
\bar{\mathbf{T}}\left[  Y^{\dagger}_+(0)Y_-(0)
\right]
|g^{K_1}(k_1)g^{K_2}(k_2) \rangle\,
&=&
\nonumber\\ && \hspace{-6.7cm }
  g^2_s\,n_-^{\eta_1}n_-^{\eta_2}\, \left( 
	\frac{1 }{n_-k_2}\,\frac{1 }{n_-(k_1+k_2)}
	\mathbf{T}^{K_2}\,\mathbf{T}^{K_1} 
	+\frac{1 }{n_-k_1}\frac{1 }{n_-(k_1+k_2)}
	\mathbf{T}^{K_1}\,\mathbf{T}^{K_2}\right)
	\epsilon_{\eta_1\,}(k_1)\epsilon_{\eta_2\,}(k_2)
\nonumber	\\ &&\hspace{-7cm} +
 g^2_s\,n_+^{\eta_1}n_+^{\eta_2}\, \left( 
	\frac{1 }{n_+k_2}\,\frac{1 }{n_+(k_1+k_2)}
	\mathbf{T}^{K_1}\,\mathbf{T}^{K_2} 
	+\frac{1 }{n_+k_1}\frac{1 }{n_+(k_1+k_2)}
	\mathbf{T}^{K_2}\,\mathbf{T}^{K_1}\right)
		\epsilon_{\eta_1\,}(k_1)\epsilon_{\eta_2\,}(k_2)
\nonumber	\\ &&\hspace{-6.5cm} +	
	 g^2_s\,\, \left( - n_-^{\eta_1}n_+^{\eta_2} 
	\frac{1 }{n_-k_1}\,\frac{1 }{n_+k_2}
	\mathbf{T}^{K_1}\,\mathbf{T}^{K_2} 
	-n_+^{\eta_1}n_-^{\eta_2}\frac{1 }{n_+k_1}\frac{1 }{n_-k_2}
	\mathbf{T}^{K_2}\,\mathbf{T}^{K_1}\right)
		\epsilon_{\eta_1\,}(k_1)\epsilon_{\eta_2\,}(k_2)
	\nonumber	\\ &&\hspace{-2cm} 	
	\,-\,\, g^2_s\,\Big(\, i\,
 f^{K_1K_2K}\,\,\mathbf{T}^{K}\,\Big)
	\,\frac{\, 1}{n_-(k_1+k_2)}
	  		 \, \frac{1}{(k_1+k_2)^2} \, 
  				\nonumber \\ &&\hspace{-6.5cm} \times
 	\Big(  -n_-^{\eta_1} (2k_1+k_2)^{\eta_2} 
  +g^{\eta_1\eta_2}\,n_-(k_1-k_2)+n_-^{\eta_2}(2k_2+k_1)^{\eta_1}
	\,\Big)\,	\epsilon_{\eta_1\,}(k_1)\epsilon_{\eta_2\,}(k_2)
	\nonumber	\\ &&\hspace{-2cm} 
  \,+\,\, g^2_s\,\Big(\, i\,f^{K_1K_2K}\,\,\mathbf{T}^{K}\,
 \Big)
	\,\frac{\, 1}{n_+(k_1+k_2)}
	  		 \, \frac{1}{(k_1+k_2)^2} \, 
  				\nonumber \\ &&\hspace{-6.5cm}\times
 	\Big(  -n_+^{\eta_1} (2k_1+k_2)^{\eta_2} 
  +g^{\eta_1\eta_2}\,n_+(k_1-k_2)+n_+^{\eta_2}(2k_2+k_1)^{\eta_1}
	\,\Big)\,	\epsilon_{\eta_1\,}(k_1)\epsilon_{\eta_2\,}(k_2),
\end{eqnarray}
\begin{eqnarray}\label{eq:LPtwoquarkab}
\langle 0| 
\bar{\mathbf{T}}\left[  Y^{\dagger}_+(0)Y_-(0)
\right]
|q (k_1)\bar{q}(k_2) \rangle\,
&=&g^2_s  \,\,\mathbf{T}^{A} \frac{1}{(k_1+k_2)^2}
\nonumber\\ && \hspace{-5.2cm }
 \times \left( -
  \frac{\, n_{+\nu} }{n_+(k_1+k_2)}
  + 
	\,\frac{\, n_{-\nu} }{n_-(k_1+k_2)}  \right)
\bar{v}(k_2)\mathbf{T}^{A}\gamma^{\nu}  u(k_1),	  		 \, \, 
  \end{eqnarray}
and
\begin{eqnarray}\label{eq:LPtwoghostkab}
\langle 0| 
\bar{\mathbf{T}}\left[  Y^{\dagger}_+(0)Y_-(0)
\right]
|{c}^{K_1} (k_1) \bar{{c}}^{K_2}(k_2) \rangle\,
&=&g^2_s i f^{K_1K_2A}\,\,\mathbf{T}^{A}
\nonumber\\ && \hspace{-6.0cm }
 \times \left( 
  \frac{\, n_+k_2 }{n_+(k_1+k_2)}
  - 
	\,\frac{\, n_-k_2 }{n_-(k_1+k_2)}\right)  
	  		 \, \frac{1}{(k_1+k_2)^2} \, 
  		 	  	 \,.
  \end{eqnarray}
For $S_3$ we need
\begin{eqnarray}\label{twogluonmatrix3bfull}
 \langle g^{K_1}(k_1)g^{K_2}(k_2)|  
 \frac{1}{(in_-\partial)^2}
 \left[\mathcal{B}^{+\,\mu_\perp}_{}(z_-),\left[in_-\partial
 \mathcal{B}^+_{\mu_\perp}(z_-) \right] \right]  
 |0 \rangle &=&
 \nonumber\\ &&\hspace{-8.0cm}
  g_s^2
  if^{K_1K_2K}\mathbf{T}^K  
 \frac{1}{(n_-(k_1+k_2))^2}
\bigg[  (
 n_-k_1 
- n_-k_2 )g_{\perp}^{\eta_1\eta_2} 
 \nonumber\\ &&
 \hspace{-8.2cm}
   +\frac{(n_-k_2)}{(n_-k_1)} k_{1\perp}^{\eta_2}n_-^{\eta_1}
  -\frac{(n_-k_1)}{(n_-k_2)}
  k_{2\perp}^{\eta_1}n_-^{\eta_2}
 +k^{\eta_1}_{2\perp}n_-^{\eta_2} 
 -k^{\eta_2}_{1\perp}n_-^{\eta_1}
  \nonumber \\ &&  \hspace{-8.2cm}
   -
 \left(\frac{k_{1\perp}\cdot k_{2\perp}}{n_-k_1}
 -\frac{k_{1\perp}\cdot k_{2\perp}}{n_-k_2}\right)
 n_-^{\eta_1}n_-^{\eta_2} \bigg]
 \epsilon^{*}_{\eta_1}(k_1)
 \epsilon^{*}_{\eta_2}(k_2)e^{iz_-(k_1+k_2)}.
  \end{eqnarray}

\section{Terms contributing to the NNLO cross section}
\label{appNNLOterms}

We report in this appendix the terms contributing to \eqref{fullNNLO}, which have been calculated in \cite{Beneke:2019oqx}. First of all one has the contribution to the partonic cross section due to the kinematic correction at NNLO:
\bea
\Delta^{kin\,(2)}_{{\rm{NLP}}}(z) &=& 
\frac{\alpha_s^2}{(4\pi)^2}\, \Bigg[ C_F^2
\bigg(\frac{16}{\epsilon^2}-\frac{192 \ln(1-z)+96}{\epsilon}+512\ln^2(1-z)
\nonumber \\  
&& 
+\, 192\ln(1-z)- 40\pi^2 -256 \bigg)  + C_F C_A \,\bigg(
\frac{88}{3\epsilon}-\frac{352\ln(1-z)}{3} \nonumber\\
&&
-\, \frac{8\pi^2}{3}+ \frac{476}{9} \bigg) + C_F n_f \bigg( 
 -\frac{16}{3\epsilon}+\frac{64\ln(1-z)}{3}-\frac{56}{9} \bigg) +\mathcal{O}(\epsilon)\Bigg]\,.
\label{eq:NNLOkin}
\eea
The terms contributing to the dynamic corrections with a collinear and a hard loop
\begin{eqnarray}\label{sig9}
\Delta^{dyn\,(2)}_{{\rm{NLP-coll}}}(z)&=& 
\frac{  \alpha_s^2 }{(4\pi)^2} 
 \Bigg[C_F^2 \,\bigg(-\frac{16}{\epsilon^2}
 +\frac{48 \ln (1-z)-20}{\epsilon } \nn \\ 
&&  -72 \ln^2(1-z)+60 \ln (1-z)+8 \pi^2-24 
  \bigg)\nonumber  \\
&& + \,C_F{C_A}
 \left( \frac{20}{\epsilon }-60 \ln (1-z)+8
   \right) +\mathcal{O}(\epsilon) \Bigg]\,,
\hspace{1.8cm}
\end{eqnarray} 
and
\begin{eqnarray}\label{sig13}
 \Delta^{dyn\,(2)}_{{\rm{NLP-hard}}}&=&  
 \frac{\alpha_s^2C^2_F}{(4\pi)^2}
    \bigg[-\frac{32}{\epsilon^3}
 +\frac{64 \ln (1-z)-16}{\epsilon^2} \nn\\
&& \hspace{0cm}+\,\frac{-64 \ln^2(1-z)+32 \ln (1-z)
 +\frac{80}{3} \left(\pi^2-3\right)}{\epsilon }
 \nonumber \\
&&\hspace{0cm}-\,\frac{8}{3} \,
 \Big(-16 \ln^3(1-z)+12 \ln^2(1-z) \nn\\
&& \hspace{0cm}
+\,20 \left(\pi^2-3\right) \ln (1-z)
-56 \zeta_3- 5 \pi^2+48\Big)
 + \mathcal{O}(\epsilon)  \bigg]\,,
\end{eqnarray} respectively.

\section{Altarelli-Parisi splitting functions} 
\label{AppAP}

We list in this appendix the Altarelli-Parisi splitting functions needed for the mass renormalisation of the bare partonic cross section given in Sec.~\ref{sec:ValidationA}. The one-loop splitting function $P_{qq}^0$ reads 
\be
P_{qq}^0(z) = 4C_F \bigg\{2 \bigg[\frac{1}{1-z}\bigg]_+ 
-1 -z + \frac{3}{2} \delta(1-z)\bigg\}.
\ee
The two-loop non-singlet splitting function $P_{qq}^{1,\rm NS}$ reads
\bea \nonumber
P_{qq}^{1,\rm NS}(z) &=& 
n_f C_F \bigg\{ \delta(1-z) 
\bigg[-\frac{2}{3} - \frac{16}{3}\zeta_2 \bigg]
-\frac{80}{9}\bigg[\frac{1}{1-z}\bigg]_+ 
- \frac{8}{3} \frac{1+z^2}{1-z}\ln z 
- \frac{8}{9} + \frac{88}{9}z \bigg\} \\ \nonumber
&&+\,  C_F^2 \bigg\{ \delta(1-z) 
\Big[3 - 24 \zeta_2 + 48 \zeta_3\Big]
- 16 \frac{1+z^2}{1-z}\ln z \ln(1-z) \\ \nonumber
&&\hspace{2.0cm}-\, 
4(1+z)\ln^2 z - 8 \bigg(2z 
+ \frac{3}{1-z}\bigg)\ln z - 40 (1-z) \bigg\} 
\\ \nonumber
&&+\, C_A C_F \bigg\{ \delta(1-z) 
\bigg[\frac{17}{3} + \frac{88}{3}\zeta_2 
-24 \zeta_3\bigg]
+ \bigg(\frac{536}{9} -16\zeta_2\bigg) \bigg[\frac{1}{1-z}\bigg]_{+} \\ \nonumber
&&\hspace{2.0cm} 
+\,  4 \frac{1+z^2}{1-z}\ln^2 z 
+ 8(1+z)\zeta_2 - \frac{4}{3}
\bigg(5 + 5z - \frac{22}{1-z}\bigg)\ln z \\ 
&&\hspace{2.0cm} 
+\, \frac{4}{9}\big(53 - 187 z \big) \bigg\}.
\eea

\bibliography{NLP}

\end{appendix}

\end{document}